\documentclass[aps,prx,twocolumn,superscriptaddress,longbibliography]{revtex4}

\usepackage[english]{babel} 
\usepackage[latin1]{inputenc}
\usepackage[usenames,dvipsnames]{color}
\usepackage{graphicx}
\usepackage{amsmath,amssymb,amsfonts}
\usepackage[colorlinks=true,linkcolor=blue,urlcolor=blue,citecolor=blue]{hyperref}
\usepackage[percent]{overpic}

\begin{document}

\title{Enhancing coherent energy transfer between quantum devices via a mediator}

\author{Alba Crescente}
\affiliation{Dipartimento di Fisica, Universit\`a di Genova, Via Dodecaneso 33, 16146, Genova, Italy}
\affiliation{CNR-SPIN, Via Dodecaneso 33, 16146, Genova, Italy}
\author{Dario Ferraro}
\affiliation{Dipartimento di Fisica, Universit\`a di Genova, Via Dodecaneso 33, 16146, Genova, Italy}
\affiliation{CNR-SPIN, Via Dodecaneso 33, 16146, Genova, Italy}
\author{Matteo Carrega}
\affiliation{CNR-SPIN, Via Dodecaneso 33, 16146, Genova, Italy}
\author{Maura Sassetti}
\affiliation{Dipartimento di Fisica, Universit\`a di Genova, Via Dodecaneso 33, 16146, Genova, Italy}
\affiliation{CNR-SPIN, Via Dodecaneso 33, 16146, Genova, Italy}

\begin{abstract}
We investigate the coherent energy transfer between two quantum systems mediated by a quantum bus. In particular, we consider the energy transfer process between two qubits, and how it can be influenced by using a third qubit or photons in a resonant cavity as mediators. Inspecting different figures of merit and considering both on and off-resonance configurations, we characterize the energy transfer performances. We show that, while the qubit-mediated transfer shows no advantages with respect to a direct coupling case, the cavity-mediated one is progressively more and more efficient as function of the number of photons stored in the cavity that acts as a quantum bus. The speeding-up of the energy transfer time, due to a quantum mediator paves the way for new architecture designs in quantum technologies and energy based quantum logics.
\end{abstract}

\maketitle

\section{Introduction}
The last decades have witnessed fast developments of quantum technologies, which are assuming a central role for a progressively broader scientific community worldwide~\cite{Riedel17,Acin18,Zhang19,Raymer19,Sussman19}. Closely related to the basic aspect of this branch of research is the growing interest in the field of quantum thermodynamics~\cite{Esposito09,Vinjanampathy16,Campisi16,Benenti17,Campisi17,Campaiolibook,Paolucci18,Bera19,Carrega19,Vischi19}, a very active topic where classical notions such as work and heat are reconsidered with the aim of characterizing the functioning of thermal machines and batteries based on, possibly out of equilibrium, quantum systems~\cite{Bhattacharjee21}. These represent highly non trivial fundamental issues, that can both explain the behaviour of quantum devices at cryogenic temperature and influence the engineering of novel architectures. For what it concerns the Quantum Batteries (QBs), starting from the seminal ideas introduced in Ref.~\cite{Alicki13}, various theoretical proposals have been elaborated with the aim of realizing miniaturized devices able to exploit genuine quantum features to store and release energy in a controlled way.  They can be implemented in set-ups conventionally used for quantum computation~\cite{Binder15,Campaioli17,Gyhm22}, in artificial atoms~\cite{Le18,Liu19,Zhang19b,Tabesh20,Rosa20,Rossini20,Crescente20,Carrega20,Seah21, Caravelli21, Mitchison21, Konar21} and in the framework of cavity and circuit quantum electrodynamics~\cite{Ferraro18,Ferraro19,Crescente20b,Delmonte21,Dou21}. These theoretical investigations represent a change of paradigm in the field of energy storage with respect to two centuries old electrochemical principles which are still at the core of nowadays technology. Remarkably, the first experimental realization of  QBs have been reported in 2021 using a collection of fluorescent organic molecules embedded in a microcavity~\cite{Quach20}. Even more recently another experiment characterizing a QB, realized with a three level superconducting qubit in the transmon regime, has been carried out~\cite{Hu21}. The possibility to simulate the behavior of a QB in the controlled environment offered by the cloud-based IBM quantum machines has been also recently investigated, showing that these kind of devices, without any ad hoc optimization, are able to compete with the performances of state of the art set-ups~\cite{Gemme22}. This testifies the great interest about the possibility to achieve fast and efficient energy storage at the quantum level.   

To date, the research on QBs has been devoted mainly to find efficient ways to store energy into quantum system and release it on demand to locally supply energy to miniaturized devices~\cite{Le18, Andolina18, Ferraro18, Crescente20, Carrega20,Centrone21}. An interesting and still largely unexplored new development is related to the possibility of coherently transfer energy among distant quantum systems, realizing the energetic counterpart of the two-qubit SWAP logic gate which plays a major role in quantum information and quantum computation~\cite{Nielsen_Book}. Remarkably enough, due to the fact that the energy stored into a QB only depends on the populations of the quantum states~\cite{Crescente20},  this "energy SWAP" should be more robust with respect to its information counterpart, being affected mainly by relaxation and only marginally by decoherence~\cite{Harochebook, Weissbook}. Moreover, the realization of this kind of process represents a crucial step towards the creation of a capillary energy networks able to connect distant parts of a fully quantum device~\cite{Scarlino19, Landig19}. 

This work fits in this growing field, aiming at characterizing the coherent energy transfer between two quantum systems. We will focus on the simple, but experimentally relevant~\cite{Dicarlo09, Lu12}, situation of two two-level systems (TLSs). The energy transfer between them will be mediated by another simple quantum system which play the role of a quantum bus. The two systems, exchanging energy through the mediator, can be seen both as a QB and its charger~\cite{Andolina18,Qi21} or as a QB and an active user of the energy stored in the battery itself. In our analysis, the role of mediator will be played by an additional TLS or by the photons confined into a resonant cavity~\cite{Lu12}. While the former architecture has been recently considered to realize high-fidelity two-qubit gates~\cite{Sung21}, the latter is routinely used for instance for the readout of superconducting qubits~\cite{Krantz19}. Moreover, the possibility to connect qubits of different nature by coupling them to the same resonant cavity has been experimentally demonstrated very recently~\cite{Scarlino19,Landig19} and represents a crucial step in the roadmap toward the full accomplishment of the second quantum revolution~\cite{Laucht21}. Interestingly, a mediated coupling can strongly increase the range of interaction between the connected devices from hundred nanometers as in the direct capacitive or inductive coupling between superconducting qubits~\cite{Krantz19} up to some centimeters when a resonant cavity plays the role of the quantum bus~\cite{Sillanpaa07,Majer07}. This opened new perspectives in the domain of quantum technologies, leading to the possibility to transfer quantum information over macroscopic lengths. 

In this paper, we characterize in detail the coherent energy transfer between two TLSs in presence of a mediator. We investigate the cases where the TLSs are both on-resonance and off-resonance, making comparison between the mediated cases and the direct coupling case, chosen as a reference benchmark.
 To this end, we will introduce and evaluate relevant figures of merit such as the energy transfer time, the energy stored in each part of the device, the work required to realize the energy transfer protocol and the overlap between the final state of the system and an optimal reference target state. According to this we determine the configuration in which the first maximum of the energy stored in the QB is achieved. We will show that the cavity-mediated transfer, in addition to lead to a longer range coupling with relevant impact on actual experimental implementations, is characterized by a progressively faster and more efficient energy transfer by increasing the number of photons trapped into the cavity. Moreover, we will observe that in this case the mediator is not only able to guarantee a complete energy transfer in the resonant case, but can also play the role of a facilitator which increases the efficiency of the energy transfer off-resonance.

The paper is organized as follows. In Section~\ref{sec:model} we introduce the model for a direct coupling between TLSs as well as the ones for TLSs interacting through an additional TLS or the photons in a resonant cavity. The more relevant figures of merit to characterize the efficiency of the devices are discussed in Section~\ref{sec:figures}. 
The main results and the comparisons of the performances of the different addressed models both on and off-resonance are reported in~\ref{sec:results}. Section~\ref{sec:conclusions} is devoted to the conclusions. Finally technical details of the calculations are discussed in three Appendices. 

\section{Model \label{sec:model}}

\begin{figure}[h!]
\centering
\includegraphics[scale=0.35]{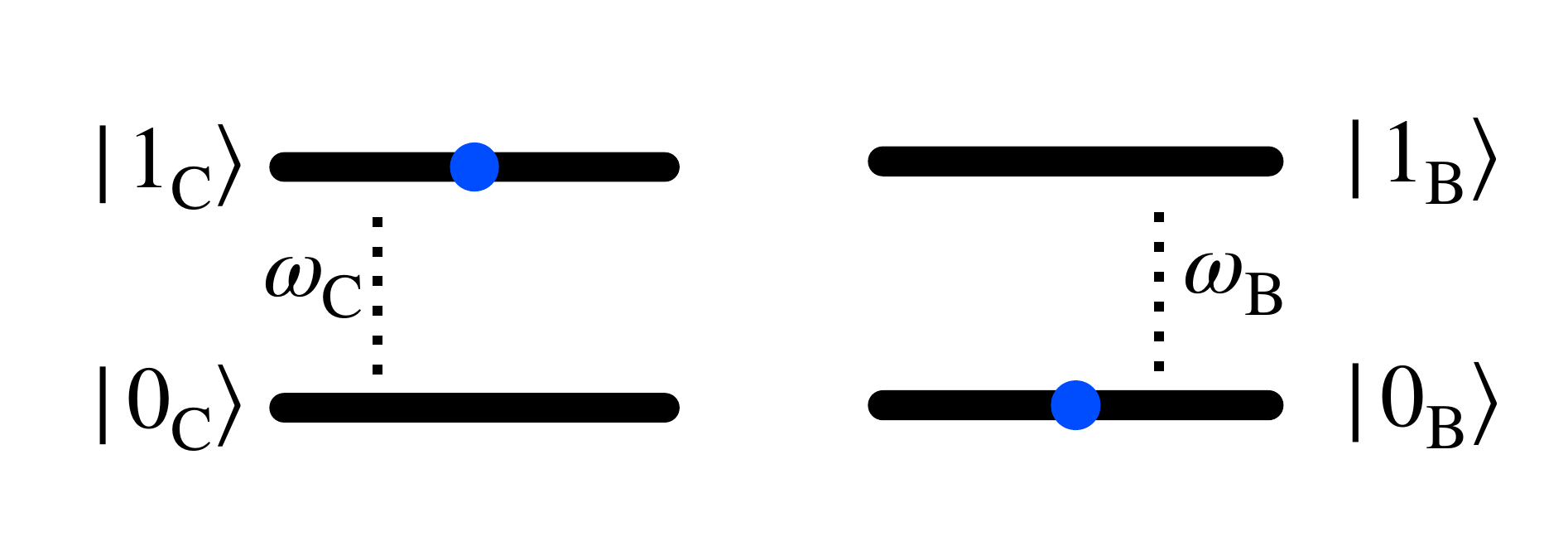} 
   \caption {Schematic representation of the two quantum systems. The first one, with energy separation $\omega_{\rm C}$, acts as a charger (C), while the second, with energy separation $\omega_{\rm B}$, can be seen as a QB (B).}
   \label{fig1}
\end{figure}

We want to investigate the energy transfer between two quantum systems. To fix the ideas, one can consider the first as a quantum charger (C) while the second as a QB (B). However, other possible configurations can be described in an analogous way. To keep the analysis as simple as possible, but still describing experimentally relevant situations, the two quantum systems are modeled as two TLSs with energy separation between the corresponding ground states $|0_{\rm C,B}\rangle$ and the excited states $|1_{\rm C,B}\rangle$ given by $\omega_{\rm C}$ and $\omega_{\rm B}$ respectively (see Fig.~\ref{fig1}).

The free Hamiltonian of this composite system is (hereafter we set $\hbar=1$)

\begin{equation} 
H_0= H_{\rm C}+H_{\rm B}=\frac{\omega_{\rm C}}{2}\sigma_z^{(\rm C)}+\frac{\omega_{\rm B}}{2}\sigma_z^{(\rm B)}, 
\label{H_0_d}
\end{equation}

\noindent where $\sigma_z^{(i)}$ is the Pauli matrix along the $\hat{z}$ direction, acting on the $i={\rm C,B}$ space. In the following we will describe different protocols that can produce energy transfer between these two entities.
Notice that, in our discussion we will consider the composite system as a closed quantum system, meaning that dissipative effects related to relaxation and dephasing phenomena are not taken into account. This is possible when the typical relaxation $t_r$ and dephasing $t_\varphi$ times are longer with respect to the considered evolution time $t$, i.e. $t_r,t_\varphi \gg t$~\cite{Devoret13, Wendin17}. 

\subsection{Direct coupling}
 Under the assumption of a local (short range) and direct capacitive coupling between the TLSs we can consider the following interaction Hamiltonian 

\begin{equation} 
H_{\rm int}^{(\rm d)}(t)= gf(t)[\sigma_-^{(\rm C)}\sigma_+^{(\rm B)}+\sigma_+^{(\rm C)}\sigma_-^{(\rm B)}], 
\label{H_int_d}
\end{equation}

\noindent where the apex $(\rm d)$ stands for direct interactions, $g$ is a coupling constant and $f(t)$ is a dimensionless time dependent function which has been introduced in order to take into account the switching on and off of the interaction. Its precise shape will be specified later.

 Here, we have defined spin ladder operators $\sigma^{(i)}_\pm=(\sigma^{(i)}_x\pm i \sigma^{(i)}_y)/2$, with $\sigma^{(i)}_{x,y}$ the Pauli matrices along the $\hat{x}, \hat{y}$ direction respectively.
 The above interaction Hamiltonian is written in the so-called rotating wave approximation (RWA)~\cite{Schweber67, Graham84, Schleich_Book} and can be derived from a capacitive coupling between superconducting qubits realizing the TLSs (see Appendix~\ref{sec:qubitmodels1} for more details). The common choice of working in RWA leads to simplification in the analysis, but imposes a constraint on admissible values for the coupling constant, namely $g \lesssim 0.1\omega_{\rm C,B}$. However, this doesn't represent a major limitation for our study due to the fact that most of the experimental realizations of such quantum systems well fits into this regime~\cite{Majer05, Niskanen07}. 

According to the above considerations, the Hamiltonian for a direct energy transfer between the charger and the QB can be written as

\begin{equation}
\label{dint} 
H_{\rm TLS}^{(\rm d)}(t)= H_0+ H_{\rm int}^{(\rm d)}(t). \end{equation}

\subsection{Mediated coupling}
\begin{figure}[h!]
\centering
\includegraphics[scale=0.38]{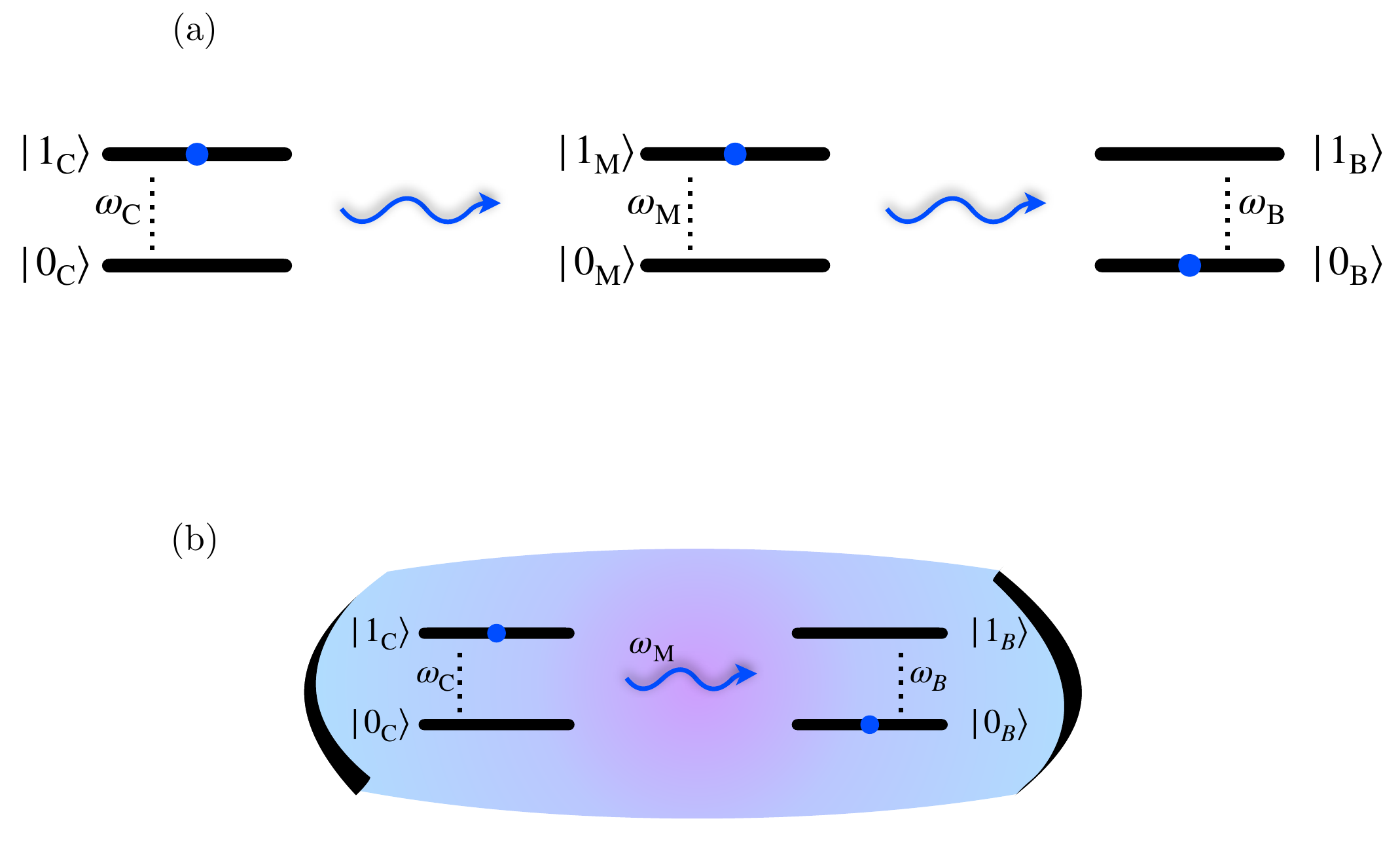} 
   \caption {Schematic representation of two possible quantum bus schemes that mediate energy transfer between two TLSs. Panel (a) represents the TLS-mediated scenario, where a third TLS with energy separation $\omega_{\rm M}$ is locally coupled to both C and B. Panel (b) represents the cavity-mediated one, where the photons of the cavity with frequency $\omega_{\rm M}$ mediates the energy transfer.}
   \label{fig2}
\end{figure}

The aim here is to compare the performances of the energy transfer for the direct coupling introduced above, with the ones where a quantum bus acts as a mediator between the two TLSs. We will focus on two possible scenarios: in the first, a third TLS allows the transfer [see Fig.~\ref{fig2} (a)], while in the second the photons of a resonant cavity play the role of mediators for the energy transfer [see Fig.~\ref{fig2} (b)].

\paragraph{TLS-mediated model.} We consider the system in Fig.~\ref{fig2} (a), where the energy transfer between C and B is mediated by a third TLS (M), with energy separation $\omega_{\rm M}$. Extending what discussed for the direct coupling case and working again in the RWA (see Appendix~\ref{appA} for more details), this model is described by the following Hamiltonian
\begin{equation}
\label{H3tls} 
H_{\rm TLS}^{(\rm m)}(t)=H_{\rm C}+H_{\rm B}+\frac{\omega_{\rm M}}{2}\sigma_z^{(\rm M)}+H^{(\rm m)}_{\rm int, TLS}(t)
\end{equation}
where
\begin{eqnarray} 
H^{(\rm m)}_{\rm int, TLS}(t)&=&g_{\rm CM}f(t)[\sigma_-^{(\rm C)}\sigma_+^{(\rm M)}+\sigma_+^{(\rm C)}\sigma_-^{(\rm M)}] \nonumber \\
&+&g_{\rm BM}f(t)[\sigma_-^{(\rm B)}\sigma_+^{(\rm M)}+\sigma_+^{(\rm B)}\sigma_-^{(\rm M)}],
\end{eqnarray}

\noindent with the apex $(\rm m)$ indicating the mediated interaction. Here $g_{\rm CM}$ and $g_{\rm BM}$ are the local coupling constants between C and M and between B and M respectively and no direct interaction between C and B is allowed. In the above expression, it appears the function $f(t)$ whose functional form is assumed to be the same as in the direct case.

\paragraph{Cavity-mediated model.} In the second scenario, depicted in Fig.~\ref{fig2} (b), the transfer process is mediated by the photons confined into a resonant cavity of characteristic frequency $\omega_{\rm M}$. The Hamiltonian of this system is

\begin{equation}
\label{Hcav} 
H_{\rm cavity}^{(\rm m)}(t)=H_{\rm C}+H_{\rm B}+\omega_{\rm M}a^\dagger a +H^{(\rm m)}_{\rm int, cavity}(t), \end{equation}
where
\begin{eqnarray}
H_{\rm int, cavity}^{(\rm m)}(t)&=&g_{\rm CM}f(t)[a^\dagger \sigma_-^{(\rm C)}+a \sigma_+^{(\rm C)}] \nonumber \\
&+&g_{\rm BM}f(t)[a^\dagger \sigma_-^{(\rm B)}+a \sigma_+^{(\rm B)}]. \end{eqnarray}

\noindent Here, $a$ ($a^\dagger$) is the annihilation (creation) operator of the photons and, as before, $g_{\rm CM}$ and $g_{\rm BM}$ are the local coupling constants between C and M (here represented by the photons in the cavity) and between B and M, respectively and again $f(t)$ is the same time dependent function introduced above. Also in this case we work in the RWA, which leads to the usual Jaynes-Cummings form of the interaction~\cite{Schleich_Book, JC63}. This kind of coupling, and the RWA scheme, can be traced back to the capacitive coupling between superconducting circuits as clarified in Appendix~\ref{appA}.

Before closing this section, we mention that we will investigate both the resonant regime ($\omega_{\rm C}=\omega_{\rm M}=\omega_{\rm B}$) and the off-resonance regime. In particular we will focus on the case 

\begin{equation}
\label{offR} 
\omega_{\rm C}=\omega_{\rm M}=\alpha\omega_{\rm B}, \end{equation}

\noindent where $\alpha$ is a positive real parameter. 

The motivation of considering off-resonance conditions is due to the difficulty of realizing absolutely identical TLSs from the experimental point of view. Therefore, following what reported in literature~\cite{Sillanpaa07, Scarlino19}, we will consider mismatches in the level spacing in the range $\alpha= 0.8 \div 1.2$. Notice that, due to the symmetries of the considered models, values of $\alpha>1$ can be obtained starting from the $\alpha<1$ case. Therefore, in the following we will mainly address this latter case. Moreover, due to the fact that the results does not change qualitatively varying the interaction constant between the systems, we will focus on the case $g_{\rm CM}=g_{\rm BM}=g$, with the same value of the coupling for both the mediated interaction cases.

\section{Figures of merit \label{sec:figures}}

To characterize the energy transfer between the two TLSs we need to study how much of the energy stored in C can be transferred to B and how fast can be this process. In addition, we will take into account the switching on and off of the interaction terms by evaluating the total work that should be  supplied in doing such operations. 

\subsection{Stored energy and charging time}
\label{Sec_IIIA}
First of all, we consider the energy stored inside C, B and M (if present). At time $t$ this is given by~\cite{Andolina18, Ferraro18}

\begin{equation}
\label{entls} E_{ i}(t)\equiv{\rm Tr}\{\rho(t) H_{i}\}-{\rm Tr}\{\rho(0) H_{i}\}, 
\end{equation}
\noindent where $i=\rm{C,B,M}$. Here, with $H_{\rm M}$ we generally indicate the Hamiltonian contribution associated to the two considered mediators. ${\rm Tr}\{\dots \}$ represents the conventional trace operation, $\rho(0)=|\psi(0)\rangle \langle \psi(0)|$ is the total density matrix of the system associated to the initial state at time $t=0$, namely  
\begin{equation} 
|\psi(0)\rangle \equiv |1_{\rm C}, 0_{\rm B}\rangle, 
\end{equation}
in the direct coupling case and 
\begin{equation} 
|\psi(0)\rangle \equiv |1_{\rm C}, 0_{\rm B}, {\rm I}_{\rm M}\rangle
\end{equation} 
in the mediated case, and $\rho(t)$ is the corresponding density matrix operator evolved in time $t$ according to the proper total Hamiltonian ($H_{\rm TLS}^{(\rm d)}$, $H_{\rm TLS}^{(\rm m)}$ or $H_{\rm cavity}^{(\rm m)}$).

In analogy to what done in Eq.~(\ref{entls}) it is also useful to define the energy associated to the interaction contributions, namely 
\begin{equation}
E_{\mathrm{int}}(t)={\rm Tr}\{\rho(t) H_{\mathrm{int}}(t)\}-{\rm Tr}\{\rho(0) H_{\mathrm{int}}(0)\}.
\label{E_int}
\end{equation}

Notice that, throughout the paper, the model of the system won't be indicated with subscripts (TLS, 3TLS, cavity) referring to the state of the system for sake of notational convenience.
Here, we assume that C starts in the excited state $|1_{\rm C}\rangle$ (the charger system is full), while B is initially in the ground state $|0_{\rm B}\rangle$ (empty battery). Concerning the state of M, where present, we will consider different possible initial states, for the moment  generically indicated by $|{\rm I}_{\rm M}\rangle$, satisfying the condition 
\begin{equation}
{\rm Tr}\{\rho(0) H_{\mathrm{int}}(0)\}=0  
\end{equation}
and thus leading to a further simplification of Eq.~(\ref{E_int}).

Moreover, we denote with 
\begin{equation} 
\label{Emax} E_{\rm B, \rm max} \equiv E_{\rm B}(t_{\rm B, \rm max}),
\end{equation}
the first local maximum achievable value of the energy stored in the QB, which occurs at the shorter charging time $t_{\rm B, \rm max}$ and with 
\begin{equation} 
\label{Emax} E_{\rm C, \rm max} \equiv E_{\rm C}(t_{\rm B, \rm max}),
\end{equation}
the value of the energy in the charger at the same time. Indeed, as we will show below, while at resonance all the maxima are equal, out of resonance this could not be the case. However, the first achieved local maximum is typically characterized by both large stored energy and average charging power~\cite{Andolina18,Ferraro18} with potentially relevant implications from the applicative point of view.

\subsection{Average work}

To fully characterize energy transfer processes, where the interaction between the systems is time-dependent, it is important to consider the power, defined as 
\begin{eqnarray} 
\label{Ptrace}
P(t)&\equiv& \frac{d}{dt}\left[{\rm Tr}\{\rho(t) H(t)\}\right]\nonumber\\ 
&=&{\rm Tr}\bigg\{\rho(t) \frac{\partial H(t)}{\partial t}\bigg\}\nonumber \\
&=&{\rm Tr}\bigg\{\rho(t) \frac{\partial H_{\mathrm{int}}(t)}{\partial t}\bigg\}.
\end{eqnarray}
In the first line of the derivation we have considered the fact that, for a closed system, the heat exchanged with the environment is zero and the variation of the total internal energy of the system is only due to the work done on it~\cite{Allahverdyan04}. In the second line we have taken into account the fact that the density matrix $\rho(t)$ evolves in time according to the total Hamiltonian $H(t)$. Finally in the last line we have made explicit the fact that only the interaction Hamiltonian parametrically depends on time.

We remark that here $H(t)$ generally indicates the Hamiltonian for the considered cases ($H_{\rm TLS}^{(\rm d)}$, $H_{\rm TLS}^{(\rm m)}$ or $H_{\rm cavity}^{(\rm m)}$).
Then the average work $W(t)$ at a given time $t$ is then given by

\begin{equation} 
W(t)=\int_0^t dt'P(t').
\label{W_t_gen}
\end{equation}

With the above definitions, we can consider the average work done to transfer energy in the various configurations. In all the considered cases the power can be written as (see Appendix~\ref{sec:dimwork} for more details).
\begin{equation} 
P(t)=(1-\alpha)\frac{d E_{\mathrm{B}}}{dt}+\frac{d E_{\mathrm{int}}}{dt}.
 \end{equation}
According to the previously discussed initial conditions, the corresponding work is then 
\begin{equation}
\label{Work} 
W(t)=(1-\alpha)E_{\rm B}(t)+E_{\rm int}(t). 
\end{equation}

\begin{figure}[h!]
\centering
\includegraphics[scale=0.50]{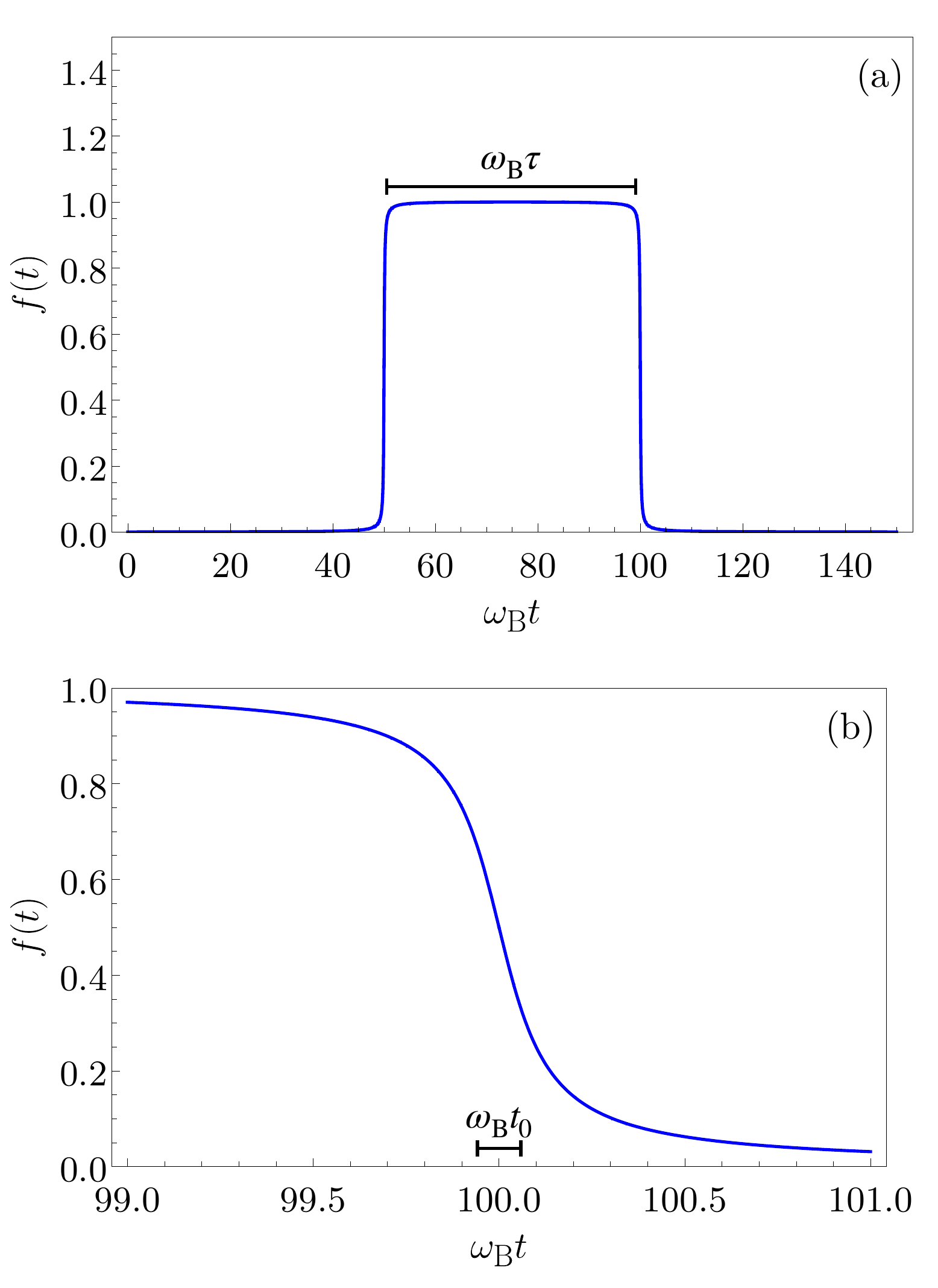} 
   \caption {(a) Behavior of $f(t)$ as function of $\omega_{\rm B}t$ for $\omega_{\rm B}\tau=50$ and $\omega_{\rm B}t_0=0.1$. (b) Zoom of the same function in correspondence of the switching off region.} 
   \label{fig2b}
\end{figure}

In order to proceed further, we need to specify the form of the function $f(t)$. From now on, we will consider the following functional form, sketched in Fig.~\ref{fig2b}, 
\begin{equation} 
f(t)= \frac{\arctan\bigg(\dfrac{t-\tau}{t_0}\bigg)-\arctan\bigg(\dfrac{t-2\tau}{t_0}\bigg)}{2\arctan\bigg(\dfrac{\tau}{2t_0}\bigg)}
\label{f_t}
\end{equation}
\noindent which describes a smooth switching on and off of the interaction between the two quantum systems C and B or between C and M and M and B in the mediated cases.
As we can see from Fig.~\ref{fig2b} the parameter $\tau$ controls the time window where the interaction is active, while $t_0$ is the width of the switching ramp. We want to underline that, despite being convenient from the numerical point of view, the chosen function isn't strictly zero at $t=0$. However, by properly setting the parameters $\tau$ and $t_0$ in such a way that $\tau\gg t_0$, its contribution at times $t\leq 0$ is negligible for all practical purposes. We can then assume a free dynamics of our system at $t=0$.
By controlling the parameters $\tau$ and $t_0$ it is then possible to turn off the interaction when the first maximum of the energy stored in B is achieved: at time $t_{\rm B, max}$ the interaction Hamiltonian $H_{\rm int}(t)$ is switched off. 
Notice that other possible smooth step-wise forms of the drive can be considered, leading to qualitatively similar results as long as $t_{0}$ is the shorter time scale involved in the process.

From the experimental point of view this time dependent modulation of the coupling could be achieved for example by changing in time the capacitance connecting the various part of the superconducting circuits discussed in Appendix~\ref{appA}. Alternative protocols involving the modulation of the Josephson energy, realized for example by replacing a single junction with a SQUID controlled in time, should lead to an analogous behaviour (see Ref.~\cite{Ferraro18} and its supporting material for a related discussion).

According to the protocol discussed above, it is instructive to inspect a relation for the average work at times such that $t^\star> t_{\rm B,max}$ where the battery has been charged and the interaction Hamiltonian has been switched off according to the considered protocol. In this regime the previous relation in Eq.~(\ref{Work}) becomes 

\begin{equation}
\label{Wstar} 
W^\star\equiv W(t^\star)=(1-\alpha)E_{\rm B,max}, 
\end{equation}

\noindent meaning that the work done to switch off the interaction is constant and it only depends on how much the target system is off-resonance  with respect to the  original one (quantified by $\alpha$) and on the maximum value of the energy stored into B ($E_{\rm B,max}$). Notice that in the resonant case one has $\alpha=1$ and then $W^\star=0$~\cite{Andolina18}.

\subsection{State SWAP quantifier}
Here, we introduce a quantifier for the efficiency of the energy SWAP between C and B. In the optimal condition we want to reach a final state of the system, after the interaction is switched off, of the form

\begin{equation} 
|\psi(t_{{\rm B}, \rm max})\rangle=|0_{\rm C},1_{\rm B}, \cdot\rangle.
\end{equation}

\noindent Notice that we do not put any constraint on the final state of M. Consequently it is useful to define a target state that only takes into account the configuration of C and B, namely

\begin{equation}\label{target_state} 
|\psi_{\rm optimal}\rangle\equiv |0_{\rm C},1_{\rm B}\rangle. \end{equation}

We can then define a quantifier related to the overlap between the state reached at time $t$ and the optimal state as~\cite{Nielsen_Book}

\begin{equation}
\label{zeta} 
\zeta(t)\equiv \rm{Tr}\{\rho_{\rm{CB}}(t)\Pi_{\rm optimal}\}, \end{equation}

\noindent where $\Pi_{\rm optimal}=|\psi_{\rm optimal}\rangle\langle\psi_{\rm optimal}|$ is the projector associated to the optimal state and

\begin{equation} 
\rho_{\rm CB}(t)=\rm{Tr}_{\rm M}\{\rho(t)\},  
\end{equation}

\noindent is the reduced density matrix of the system, obtained from the full density matrix of the system $\rho(t)$ by tracing out the mediator degrees of freedom. Obviously this trace operation becomes trivial in the direct coupling case, where the mediator M is absent.

This quantifier provide information about the possibility to actually realize a SWAP in the quantum states of the two TLSs, leading to a complete transfer of the quantum state of the charger to the battery. According to the definition in Eq.~(\ref{zeta}), $\zeta(t)\leq 1$, the equality holding when a complete SWAP in the quantum state of C and B is achieved. 

\section{Results \label{sec:results}}

In this Section we report and discuss the main results for the energy transfer, comparing the various mechanisms introduced in the previous Section. 

\subsection{Comments on the exact diagonalization \label{num}}

To evaluate the figures of merit and to characterize the performances of the various configurations, we resort to exact diagonalization method and numerical solution of the Schr\"odinger equation $id|\psi(t)\rangle/dt=H(t)|\psi(t)\rangle$, where again $H(t)$ indicates the Hamiltonian of the considered set-up ($H_{\rm TLS}^{(\rm d)}$, $H_{\rm TLS}^{(\rm m)}$ or $H_{\rm cavity}^{(\rm m)}$). The state of the system $|\psi(t)\rangle$, at a given time $t$, can be written projecting on a set of basis eigenstates $|\varphi_k\rangle$ of the Hamiltonian in absence of interactions, with $k$ dimension of the subset of the Hilbert space which can be spanned according to the constraints imposed by the RWA (see Appendix~\ref{sym} for more details), in such a way that 

\begin{equation}
\label{psit} 
|\psi(t)\rangle=\sum_k c_k(t)|\varphi_k\rangle. 
\end{equation}

\noindent Here, $c_k(t)$ are time-dependent coefficients which can be found numerically inserting Eq.~(\ref{psit}) into the Schr\"odinger equation. Once $|\psi(t)\rangle$ is found, it is straightforward to derive $\rho(t)$ and the different quantities of interest.


\subsection{Direct coupling \label{sec:resultsA}}

To begin with, we discuss the direct energy transfer between TLSs, which acts as a reference for the mediated cases. We will focus on the case $g=0.05\omega_{\rm B}$ (where the RWA is well justified).
 In addition, as stated above, the charger-QB system will always be in the initial state $|\psi(0)\rangle=|1_{\rm C},0_{\rm B}\rangle$. 

In passing, we mention that in the case of piecewise constant interaction this problem can be solved analytically~\cite{Andolina18}. Moreover, as long as the switching time $t_{0}$ is the shortest time scale involved in the problem, the numerical results, derived in the time dependent coupling case, differ only slightly from the ones derived in this simple analytical model. 

\begin{widetext}

\begin{figure}[h!]
\centering
\includegraphics[scale=0.50]{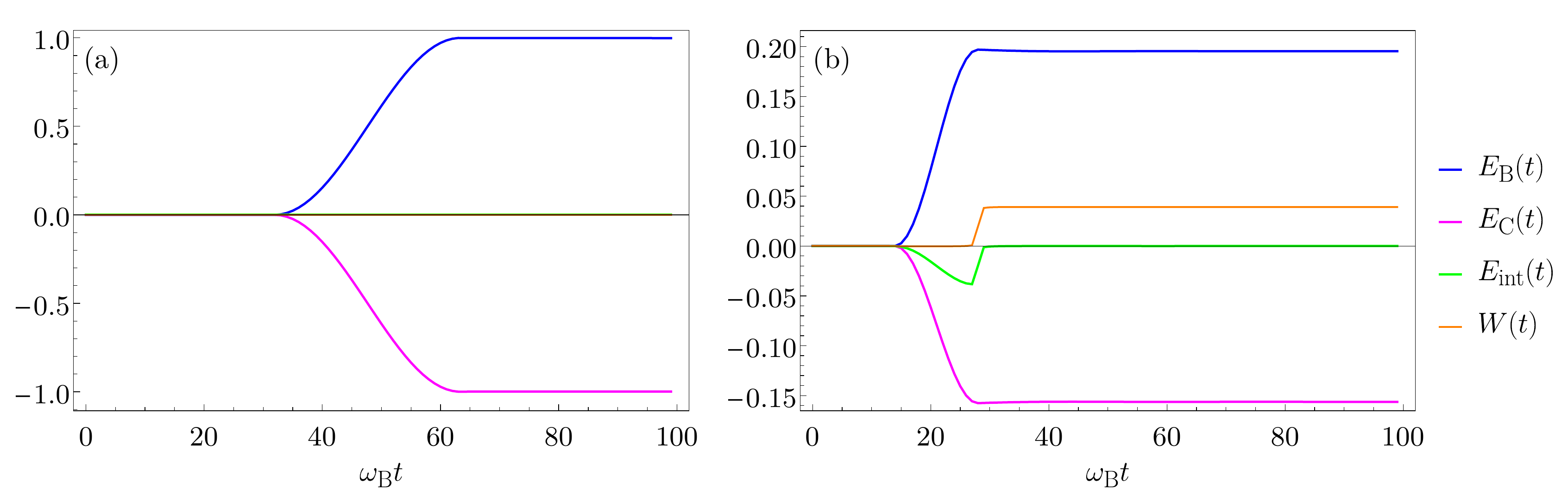} 
 \caption {Behavior (in units of $\omega_{\rm B}$) of $E_{\rm B}(t)$ (blue curves), $E_{\rm C}(t)$ (purple curves), $E_{\rm int}(t)$ (green curves) and $W(t)$ (orange curves) as function of $\omega_{\rm B}t$ in the case of direct energy transfer. Here we report the resonant $\omega_{\rm C} =\omega_{\rm B}$ (a) and off-resonant case $\omega_{\rm C}=0.8\omega_{\rm B}$ (b). The values $\omega_{\rm B}\tau=32$ and $\omega_{\rm B}\tau=14.5$ are considered respectively to switch off the interaction when the first maximum of the transferred energy is achieved. Other parameters are $g=0.05 \omega_{\rm B}$ and $\omega_{\rm B}t_0=0.1$. }
   \label{fig4}
\end{figure}

\end{widetext}

In Fig.~\ref{fig4} we report the time evolution of various figures of merit discussed in Section \ref{sec:figures} for both the resonant case (panel a) and for the off-resonance case $\alpha=0.8$ (panel b). Notice that, as stated above, we are considering only value of $\alpha<1$ since the opposite case leads to qualitative similar behavior. Moreover, from now on $\omega_{\rm B}$ will be the reference scale for all the energies. 

When the system is on-resonance we see that the energy $E_{\rm C}(t)$ stored into the charger goes from zero to $-\omega_{\rm B}$ (the negative sign indicating a reduction of energy), while $E_{\rm B}(t)$ evolve in a mirrored way from $0$ to $+\omega_{\rm B}$. During this process both the energy associated to the interaction $E_{\rm int}(t)$ and the work $W(t)$ remains zero as a consequence of the considered initial state ($|\psi(0)\rangle = |1_{\rm C},0_{\rm B}\rangle$) and of the fact that the final state reached by the system is $|\psi(t_{\rm{B, max}})\rangle = |0_{\rm C},1_{\rm B}\rangle$. These observations allow to conclude that, in this case, all the energy can be transferred directly from the charge to the QB. In particular, while the first can be completely discharged the second can achieve a perfect charging.

Different is the situation when the system is off-resonance. Here in fact, the charger loses only a fraction of its energy and the remaining part of the energy transferred to the QB is provided by the interaction term. However, even combining these two contributions the charging of the QB is limited to $\sim 20\%$. In this case the final state is not characterized by a perfect SWAP, but is given by a coherent superposition of the basis states used to numerically diagonalize the Hamiltonian $H^{\rm (d)}_{\rm TLS}$ (see Appendix~\ref{sym}). This reflects in the fact that both the energy due to the interaction and the work done to switch it off are different from zero. Moreover, we notice that, due to the fact that we are considering the evolution of a closed system, in this case one has
\begin{equation}
E_{\rm C}(t)+E_{\rm B}(t)+E_{\rm int}(t)=W(t). 
\end{equation}

In the following we will compare the above results with the ones obtained when a mediator of the energy transfer is added to the composite system. 

\subsection{TLS-mediated coupling}

We now focus on the first mediated case, where an additional TLS is introduced between C and B.

\begin{widetext}

\begin{figure}[h!]
\centering
\includegraphics[scale=0.50]{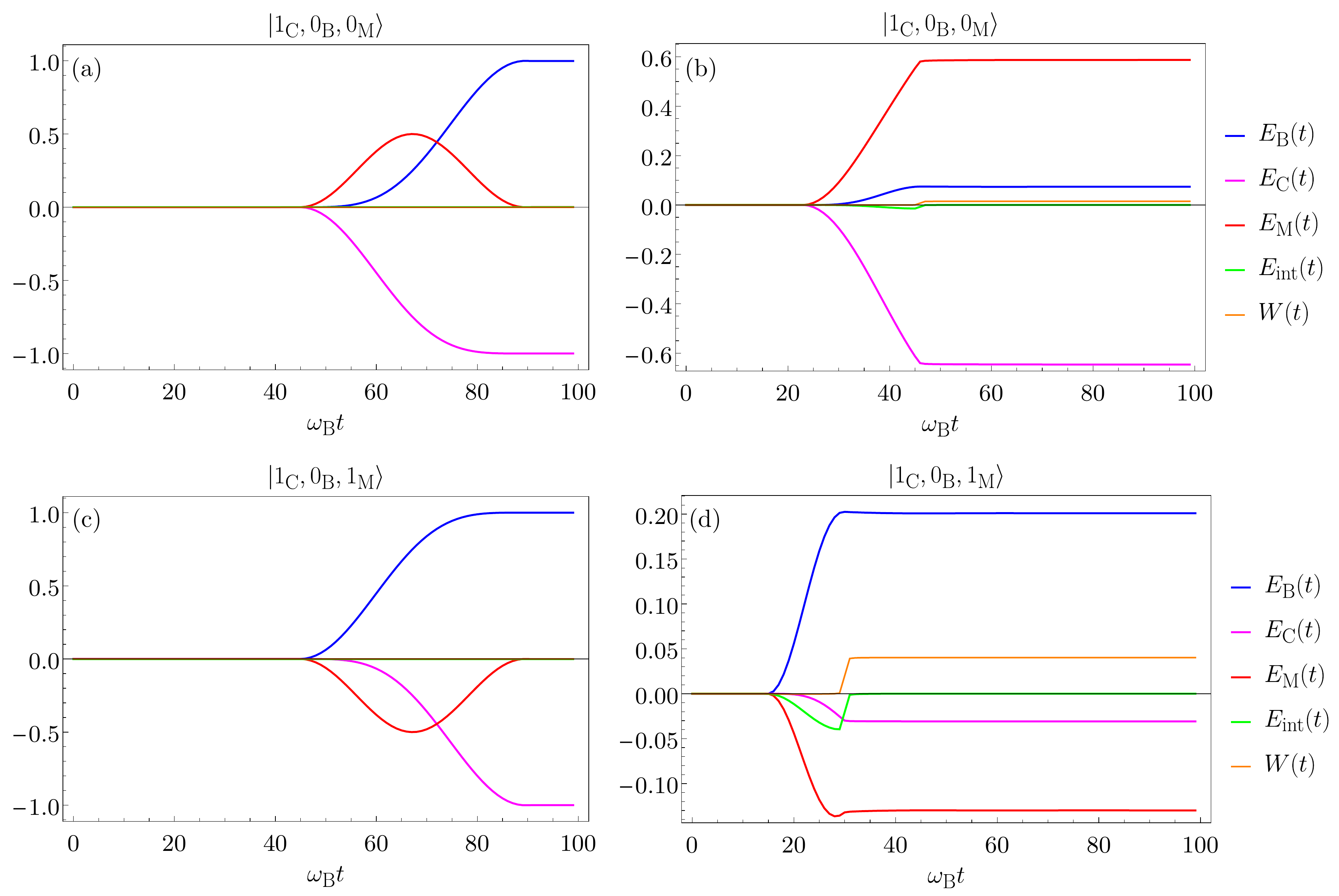} 
   \caption {Behavior (in units of $\omega_{\rm B}$) of $E_{\rm B}(t)$ (blue curves), $E_{\rm C}(t)$ (purple curves), $E_{\rm M}(t)$ (red curves), $E_{\rm int}(t)$ (green curves) and $W(t)$ (orange curves) as function of $\omega_{\rm B}t$ in the case of coupling mediated by a TLS. Here we report the resonant $\omega_{\rm C} =\omega_{\rm B}$ (a)-(c) and off-resonant case $\omega_{\rm C}=0.8\omega_{\rm B}$ (b)-(d) for different initial conditions of the mediator. The values $\omega_{\rm B}\tau=45$ (a), $\omega_{\rm B}\tau=23$ (b), $\omega_{\rm B}\tau=45$ (c)  and $\omega_{\rm B}\tau=15$ (d) are considered respectively to switch off the interaction when the first maximum of the transferred energy is achieved. Other parameters are $g=0.05 \omega_{\rm B}$ and $\omega_{\rm B}t_0=0.1$.}
   \label{fig5}
\end{figure}

\end{widetext}

In Fig.~\ref{fig5} we report the time evolution of the various figures of merit discussed in Section \ref{sec:figures} for both the resonant case (left panels) and for the off-resonance case with $\alpha=0.8$ (right panels). We also address two different initial states, namely:  $|\psi(0)\rangle=|1_{\rm C},0_{\rm B},0_{\rm M}\rangle$ (upper panels) where the mediator is in its ground state and $|\psi(0)\rangle=|1_{\rm C},0_{\rm B},1_{\rm M}\rangle$ (lower panels) where the mediator is in its excited state.

When the system is on-resonance and the mediator is initially in the ground state [see Fig.~\ref{fig5} (a)], C starts transferring energy to M. Only at a later time B can extract energy to M. Even if quite slow, this sort of bucket brigade procedure results in a complete energy transfer form C to B, in analogy with what observed in the direct coupling case, and in a SWAP of the states of C and B (see parameters reported in Table \ref{tab1}). Similarly to what discussed before this resonant condition is characterized by zero interaction energy and work. Different is the situation where the system is off-resonance and the mediator is initially in the ground state [see Fig.~\ref{fig5} (b)]. Here, C does not discharge completely and transfers only a fraction of its energy to M. However, M is not able to pass its energy to B which reaches a charge lower then $\sim 10\%$ of the maximum possible value. In this case the interaction energy, and consequently the work required to switch it off is closed to zero. Moreover, as reported in Table \ref{tab1}, the target state is almost completely missed in this case. Extending what discussed in the off-resonant direct coupling case here we have the overall constraint 
\begin{equation}
E_{\rm C}(t)+E_{\rm B}(t)+E_{\rm M}(t)+E_{\rm int}(t)=W(t). 
\label{constraint}
\end{equation}

We consider now a situation in which the mediator is initially in the excited state. When the system is on-resonance [see Fig.~\ref{fig5} (c)] B reaches a complete charging. However, in this case the energy is provided by M, whose energy is only subsequently reestablished by C. This process is characterized by a time scale identical to the one of the case in Fig.~\ref{fig5} (a) (see Table \ref{tab1}) and similarly leads to a perfect energy and state transfer from C to B with zero interaction energy and work. Conversely, off-resonance [see Fig.~\ref{fig5} (d)] M transfer its energy to B with a minor help from the interaction. However, this energy contribution is only poorly reestablished by C. This makes also this configuration not efficient from the point of view of the energy transfer (see also the other data reported in Table \ref{tab1}). As before the energy balance is given by Eq.~(\ref{constraint}).

 \begin{widetext}

\begin{table}[!h]
\centering
 \begin{tabular}{|c |c| c | c | c | }
    \hline
       \hspace{1cm} & $|\psi(0)\rangle=|1_{\rm C},0_{\rm B}\rangle$ &\hspace{1cm} $|\psi(0)\rangle=|1_{\rm C},0_{\rm B},0_{\rm M}\rangle$ \hspace{1cm} &  \hspace{1cm} $|\psi(0)\rangle=|1_{\rm C},0_{\rm B},1_{\rm M}\rangle$  \hspace{1cm}   \\ 
   \cline{2-4} &\begin{tabular}{cccc}\hspace{0.cm}$t_{\rm B, max}$&$\hspace{0.3cm} E_{\rm B,max}$ &$\hspace{0.3cm} E_{\rm C,max}$&$\hspace{0.3cm} \zeta_{\rm max}$\end{tabular}&\begin{tabular}{cccc}\hspace{0.cm}$t_{\rm B, max}$&$\hspace{0.3cm} E_{\rm B,max}$&$\hspace{0.3cm} E_{\rm C,max}$ &$\hspace{0.3cm} \zeta_{\rm max}$\end{tabular}&\begin{tabular}{cccc}\hspace{0.cm}$t_{\rm B, max}$&$\hspace{0.3cm} E_{\rm B,max}$&$\hspace{0.3cm} E_{\rm C,max}$ &$\hspace{0.3cm} \zeta_{\rm max}$\end{tabular} \\ 
   \hline
     $\alpha=1$ &\begin{tabular}{cccc} $\hspace*{0cm}64$ & $\hspace{0.6cm}\quad 1$ & $\hspace{0.7cm}\quad -1$& $\hspace{0.5cm}\quad 1$  \end{tabular} &\begin{tabular}{cccc} $\hspace{0.cm} 90$ & $\hspace{0.6cm}\quad 1$& $\hspace{0.7cm}\quad -1$ & $\hspace{0.5cm}\quad 1$\end{tabular} & \begin{tabular}{cccc} $\hspace{0.cm} 90$ & $\hspace{0.6cm}\quad 1$& $\hspace{0.7cm}\quad -1$ & $\hspace{0.5cm}\quad 1$  \end{tabular}\\ \hline 
             
     $\alpha=0.8$ &\begin{tabular}{cccc} $\hspace{0.1cm}29$ & $\hspace{0.3cm}\quad 0.197$ & $\hspace{0.cm}\quad -0.157$& $\hspace{0.cm}\quad 0.197$\end{tabular}&\begin{tabular}{cccc} $\hspace{0.1cm}46$ & $\hspace{0.3cm}\quad 0.080$ & $\hspace{0.cm}\quad -0.623$& $\hspace{0. cm}\quad 0.071$\end{tabular}&\begin{tabular}{cccc} $\hspace{0.1cm} 30$ & $\hspace{0.3cm}\quad 0.203$ & $\hspace{0.cm}\quad -0.029$& $\hspace{0.cm}\quad 0.037$\end{tabular}  \\ \hline      
    \end{tabular}
    \caption{Charging time $t_{\rm B, max}$ (in units of $\omega_{\rm B}^{-1}$), maximum of the energy in the QB $E_{\rm B, max}$ (in units of $\omega_{\rm B}$), corresponding energy in the charger $E_{\rm C, max}$ (in units of $\omega_{\rm B}$) and maximum of the overlap quantifier $\zeta_{\rm max}$ in the direct coupling (as reference benchmark) and in the TLS-mediated model, for different initial states $|\psi(0)\rangle$ and values of $\alpha$. }
    \label{tab1}
\end{table}
\end{widetext}

In passing we comment on an initial condition in which the mediator is in a superposition state [$|\psi(0)\rangle=|1_{\rm C},0_{\rm B}\rangle\otimes\frac{1}{\sqrt{2}}(|0_{\rm M}\rangle+|1_{\rm M}\rangle)$]. This is characterized by charging times and energy transfer in between the two "classical" configurations of the mediator described above (not shown). This seems to indicate the fact that, in actual experiments, there is no advantage in carefully engineering a quantum state for M in order to boost the overall performances of the device.

To summarize the results of this part, the addition of a TLS as a mediator for energy transfer processes offers no advantages in improving performances with respect to the direct coupling case.

\subsection{Cavity-mediated coupling}

\begin{widetext}

\begin{figure}[h!]
\centering
\includegraphics[scale=0.50]{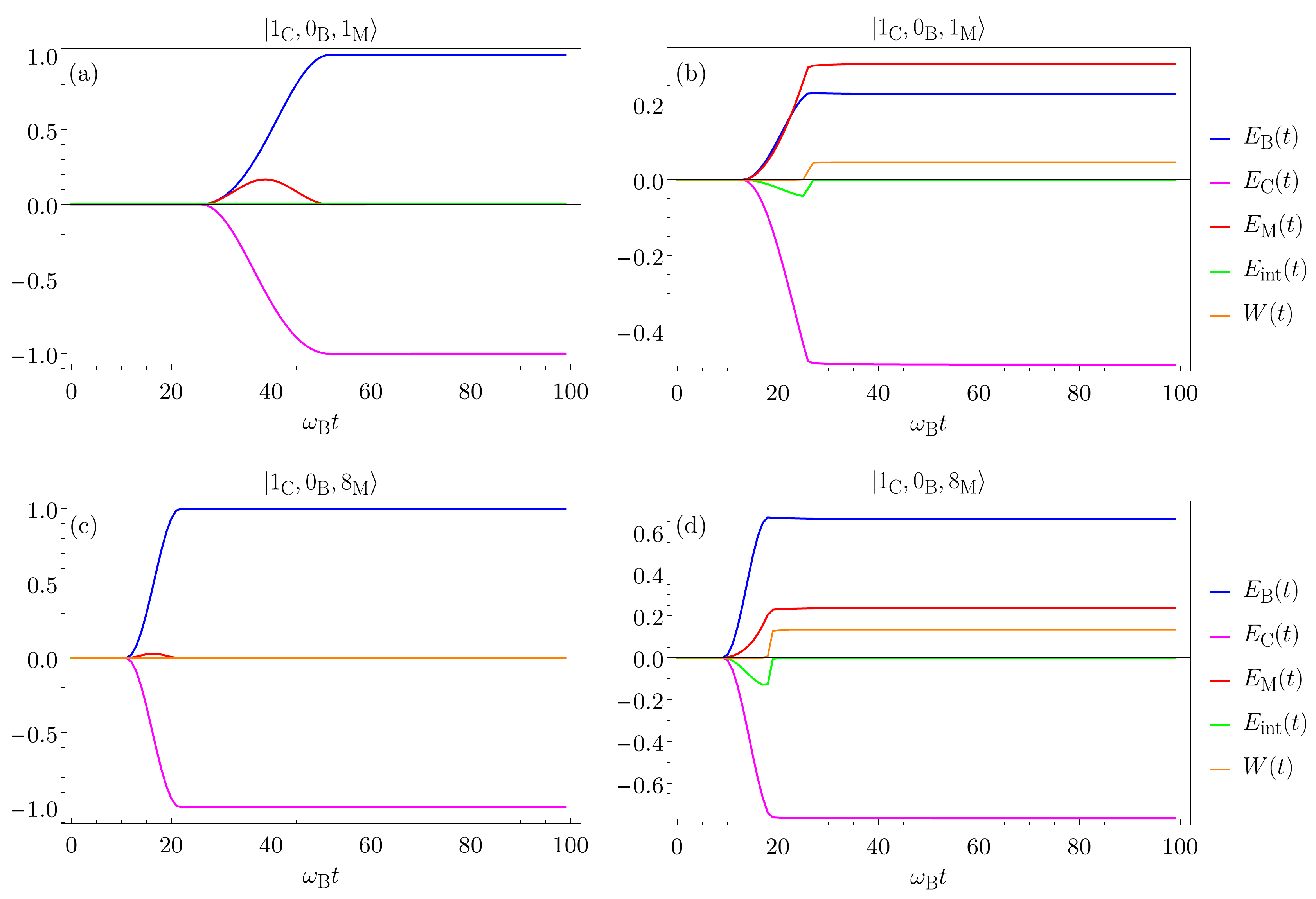} 
   \caption {Behavior (in units of $\omega_{\rm B}$) of $E_{\rm B}(t)$ (blue curves), $E_{\rm C}(t)$ (purple curves), $E_{\rm M}(t)$ (red curves), $E_{\rm int}(t)$ (green curves) and $W(t)$ (orange curves) as function of $\omega_{\rm B}t$ in the case of coupling mediated by photons trapped into a cavity. Here we report the resonant $\omega_{\rm C} =\omega_{\rm B}$ (a)-(c) and off-resonant case $\omega_{\rm C}=0.8\omega_{\rm B}$ (b)-(d) for different initial conditions of the mediator. The values $\omega_{\rm B}\tau=26.5$ (a), $\omega_{\rm B}\tau=13.5$ (b), $\omega_{\rm B}\tau=10.5$ (c)  and $\omega_{\rm B}\tau=9.5$ (d) are considered respectively to switch off the interaction when the first maximum of the transferred energy is achieved. Other parameters are $g=0.05 \omega_{\rm B}$ and $\omega_{\rm B}t_0=0.1$.}
   \label{fig6}
\end{figure}

\end{widetext}

We analyze now the case in which the energy transfer is mediated by photons in a resonant cavity. Here, it is possible to exploit the additional degree of freedom offered by the number of photons $n$ to improve the performances of the energy transfer in the composite system. As before we will start from the two possible initial states: one where there is only one photon into the cavity (upper panels of Fig.~\ref{fig6} with $|\psi(0)\rangle=|1_{\rm C},0_{\rm B},1_{\rm M}\rangle$), and a Fock state with an higher number of photons. In particular, in the following we will focus as an example on the case $n=8$ (lower panels of Fig.~\ref{fig6} with $|\psi(0)\rangle=|1_{\rm C},0_{\rm B},8_{\rm M}\rangle$). Notice that an initial state with no photons in the cavity shows the same behavior as in the case in which the interaction is mediated by an empty TLS [see Fig.~\ref{fig5} (a) and (b)]. Also in this case, we will consider both a resonant (left panels of Fig.~\ref{fig6}) and an off-resonant case with $\alpha=0.8$ (right panels of Fig.~\ref{fig6}).

When the system is on-resonance and only one photon is initially confined into the cavity (see Fig.~\ref{fig6} (a)), both the energy transfer from C to M and from M to B occurs simultaneously, leading to a complete discharging of C and charging of B in a time which is shorter with respect to both the direct coupling and the TLS mediated case (see Table \ref{tab2}). This situation is also characterized by a perfect SWAP and by zero interaction energy and work. When the system is off-resonance [see Fig.~\ref{fig6} (b)] C releases a consistent fraction of its energy. Part of it remains trapped into M, however, also thank to a contribution from the interaction, B can be charged more than $\sim 20\%$. This value is slightly greater than what observed in the direct coupling case and is achieved in a comparable time (see Table \ref{tab2}). Notice that, the overall energy balance is again constrained by Eq.~(\ref{constraint}).

More interesting is the situation in which the mediator is characterized by an higher number of photons. When the system is on-resonance [see Fig.~\ref{fig6} (c)] the behaviour is quite similar to the one discussed above for a single photon. However, the energy dynamics of M is less pronounced and the energy transfer and state SWAP from C to B occur faster (see Table \ref{tab2}). Relevant is also the impact of this richer structure of the mediator in the off-resonant case [see Fig.~\ref{fig6} (d)]. Here, C releases almost all its energy. Even if a quite important fraction of this energy remains trapped into M, it is possible, also thank to a small contribution from the interaction term, to charge B more then $\sim 60 \%$ in a very short time (see Table \ref{tab2}). In this case the mediator play the role of facilitator for the energy transfer. This represents a major improvement with respect to both the direct and the TLS-mediated coupling and could have an important impact for practical applications.

 \begin{widetext}

\begin{table}[!h]
\centering
 \begin{tabular}{|c |c| c | c | c | }
    \hline
       \hspace{1cm} & $|\psi(0)\rangle=|1_{\rm C},0_{\rm B}\rangle$ &\hspace{1cm} $|\psi(0)\rangle=|1_{\rm C},0_{\rm B},1_{\rm M}\rangle$ \hspace{1cm} &  \hspace{1cm} $|\psi(0)\rangle=|1_{\rm C},0_{\rm B},8_{\rm M}\rangle$  \hspace{1cm}   \\ 
   \cline{2-4} &\begin{tabular}{cccc}\hspace{0.cm}$t_{\rm B, max}$&$\hspace{0.3cm} E_{\rm B,max}$ &$\hspace{0.3cm} E_{\rm C,max}$&$\hspace{0.3cm} \zeta_{\rm max}$\end{tabular}&\begin{tabular}{cccc}\hspace{0.cm}$t_{\rm B, max}$&$\hspace{0.3cm} E_{\rm B,max}$&$\hspace{0.3cm} E_{\rm C,max}$ &$\hspace{0.3cm} \zeta_{\rm max}$\end{tabular}&\begin{tabular}{cccc}\hspace{0.cm}$t_{\rm B, max}$&$\hspace{0.3cm} E_{\rm B,max}$&$\hspace{0.3cm} E_{\rm C,max}$ &$\hspace{0.3cm} \zeta_{\rm max}$\end{tabular} \\ 
   \hline
     $\alpha=1$ &\begin{tabular}{cccc} $\hspace*{0cm}64$ & $\hspace{0.6cm}\quad 1$ & $\hspace{0.7cm}\quad -1$& $\hspace{0.5cm}\quad 1$  \end{tabular} &\begin{tabular}{cccc} $\hspace{0.cm}53$ & $\hspace{0.6cm}\quad 1$& $\hspace{0.7cm}\quad -1$ & $\hspace{0.5cm}\quad 1$\end{tabular} & \begin{tabular}{cccc} $\hspace{0.cm} 21$ & $\hspace{0.6cm}\quad 1$& $\hspace{0.7cm}\quad -1$ & $\hspace{0.5cm}\quad 1$  \end{tabular}\\ \hline 
             
     $\alpha=0.8$ &\begin{tabular}{cccc} $\hspace{0.1cm}29$ & $\hspace{0.3cm}\quad 0.197$ & $\hspace{0.cm}\quad -0.157$& $\hspace{0.cm}\quad 0.197$\end{tabular}&\begin{tabular}{cccc} $\hspace{0.1cm}27$ & $\hspace{0.3cm}\quad 0.237$ & $\hspace{0.cm}\quad -0.529$& $\hspace{0. cm}\quad 0.149$\end{tabular}&\begin{tabular}{cccc} $\hspace{0.1cm} 19$ & $\hspace{0.3cm}\quad 0.667$ & $\hspace{0.cm}\quad -0.780$& $\hspace{0.cm}\quad 0.587$\end{tabular}  \\ \hline      
    \end{tabular}
    \caption{Charging time $t_{\rm B, max}$ (in units of $\omega_{\rm B}^{-1}$), maximum of the energy in the QB $E_{\rm B, max}$ (in units of $\omega_{\rm B}$), corresponding energy in the charger $E_{\rm C, max}$ (in units of $\omega_{\rm B}$) and maximum of the overlap quantifier $\zeta_{\rm max}$ in the direct coupling (as reference benchmark) and in the cavity-mediated model, for different initial states $|\psi(0)\rangle$ and values of $\alpha$. }
    \label{tab2}
\end{table}
\end{widetext}

\begin{figure}[h!]
\centering
\includegraphics[scale=0.50]{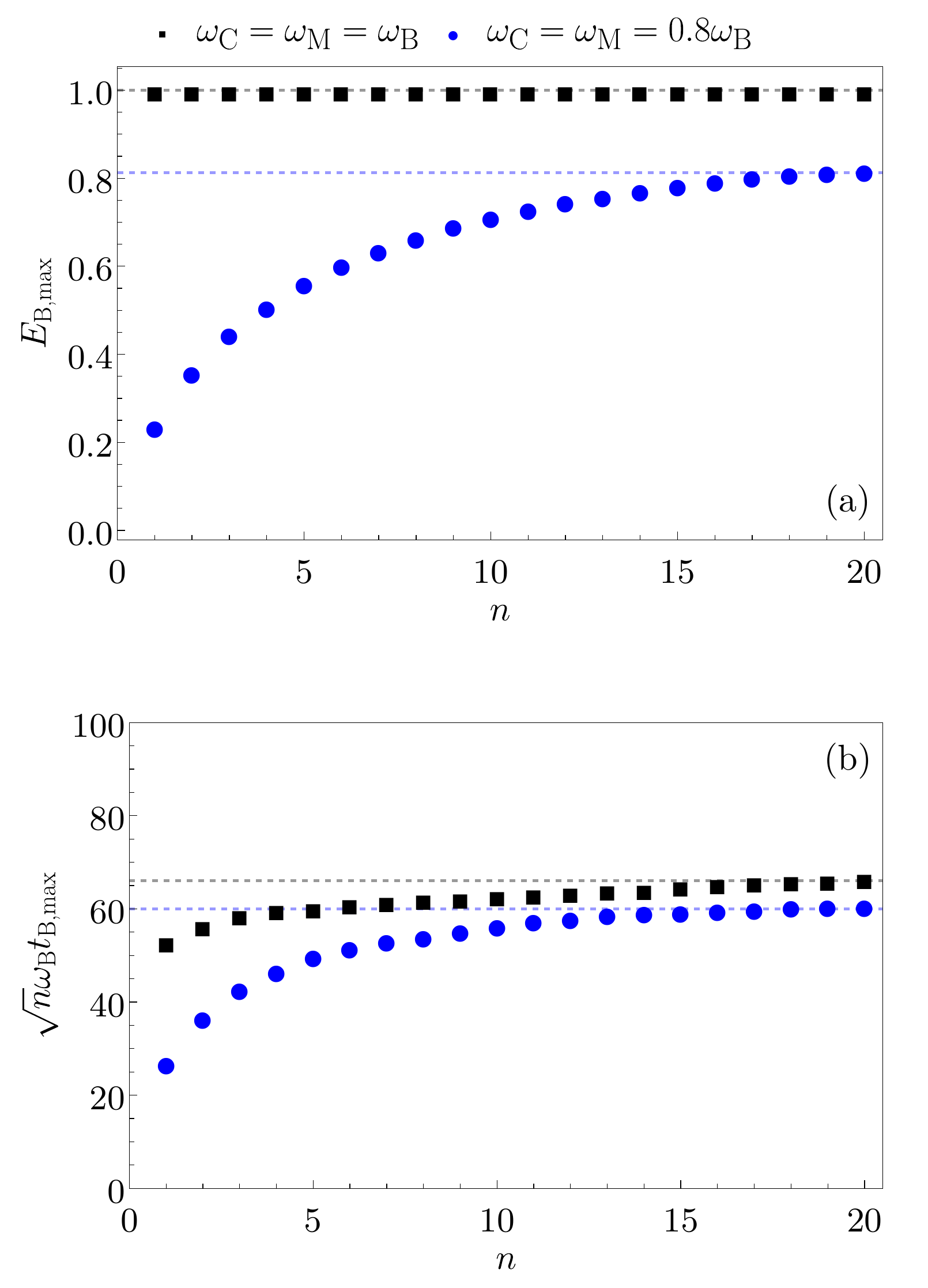} 
   \caption {(a) $E_{\rm B, max}$ (in units of $\omega_{\rm B}$) and (b) $\sqrt{n}\omega_{\rm B}t_{\rm B,max}$ (c) as function of the number of photons $n$ for the cavity-mediated model for $\omega_{\rm C}=\omega_{\rm M}=\omega_{\rm B}$ (black full squares) and $\omega_{\rm C}=\omega_{\rm M}=0.8\omega_{\rm B}$ (blue full dots). In panel (a) the large $n$ values $E_{\rm {B, max}}\sim 1.0\omega_{\rm B}$ and $E_{\rm {B, max}}\sim 0.821\omega_{\rm B}$ are represented by the black and blue dashed lines for the resonant and $\alpha=0.8$ cases respectively. In panel (b) the large $n$ value of the energy transfer time $\sqrt{n}\omega_{\rm B}t_{\rm{B, max}}\sim 66.63$ is represented by the black dashed line and $\sqrt{n}\omega_{\rm B}t_{\rm{B, max}}\sim 60.82 $ is represented by the blue dashed line.
   Other parameters are $g=0.05\omega_{\rm B}$ and $\omega_{\rm B}t_0=0.1$. Values of $\tau$ are chosen for each point in such a way to switch off the energy transfer in the system when the first maximum of the energy stored in the QB is achieved.}
   \label{fig7}
\end{figure}


These features are further enhanced by increasing the number of photons $n$ into the cavity. This can be seen from Fig.~\ref{fig7}, where the maximum of energy stored in the QB [panel (a)] and the energy transfer times [panel (b)] are reported as function of $n$. From panel (a) we can see that by increasing the number of photons in the cavity it is possible to consistently improve the energy transferred in the QB also in the off-resonant case. In fact, for $\alpha=0.8$ at large $n$ we obtain a charging of B exceeding $\sim 80 \%$, which is significantly better than the one reported in Table~\ref{tab2} for $n=8$.
The advantages in using a larger number of photons can also be seen from the charging times. Indeed, at large values of $n$ the energy transfer time scales as $t_{\rm{B, max}}\propto n^{-1/2}$ both on- and off-resonance [see Fig.~\ref{fig7} (b)].

It is worth to mention that also in this case, when the state of the cavity is a superposition between two Fock state at different $n$ (not shown) the obtained performance are in between the results for the fixed $n$ cases involved in the superposition. 
In addition, coherent states in the cavity (not shown) at average photon number $n$ are less performant with respect to the corresponding Fock states at fixed $n$ as long as this number is small. For large $n$ the performances in these two cases tend to overlap~\cite{Andolina18,Delmonte21}.

To summarize these results, we remark that there is a relevant advantage in performing energy transfer using the photons in a resonant cavity as the mediator. Indeed, we have obtained better performances concerning both the charging times and the transferred energy. In addition, the possibility to control and increase the number $n$ of photons into the cavity allows to improve the performances of the device also in the off-resonant cases, with important implications for actual experimental implementations~\cite{Sillanpaa07, Scarlino19}.

\subsection{Further comparison between the models \label{sec:summary}}

In this last part of the paper we will focus on the comparison between the models for what it concerns the maximum energy stored into the QB ($E_{\rm B,max}$), the energy extracted from the charger [$E_{\rm C,max}=E_{\rm C}(t_{\rm B, max})$] and the charging times ($t_{\rm B,max}$), as a function of the mismatch in the level spacing of the TLSs. We will consider the direct coupling model and compare it with the mediated models where the initial state of the mediator is $|1_{\rm M}\rangle$ (completely full for the TLS-mediated case and occupied by a single photon in the cavity-mediated one) and also with the case where there are $n=8$ photons in the cavity ($|8_{\rm M}\rangle$), as a representative example of multi-photon state. The plots are given as a function of the off-resonance parameter $\beta=1-\alpha$, in the range $-0.2 \div 0.2$ and show a symmetry for positive and negative $\beta$ as a consequence of the connections between $\alpha<1$ and $\alpha>1$ mentioned above.

\begin{figure}[h!]
\centering
\includegraphics[scale=0.50]{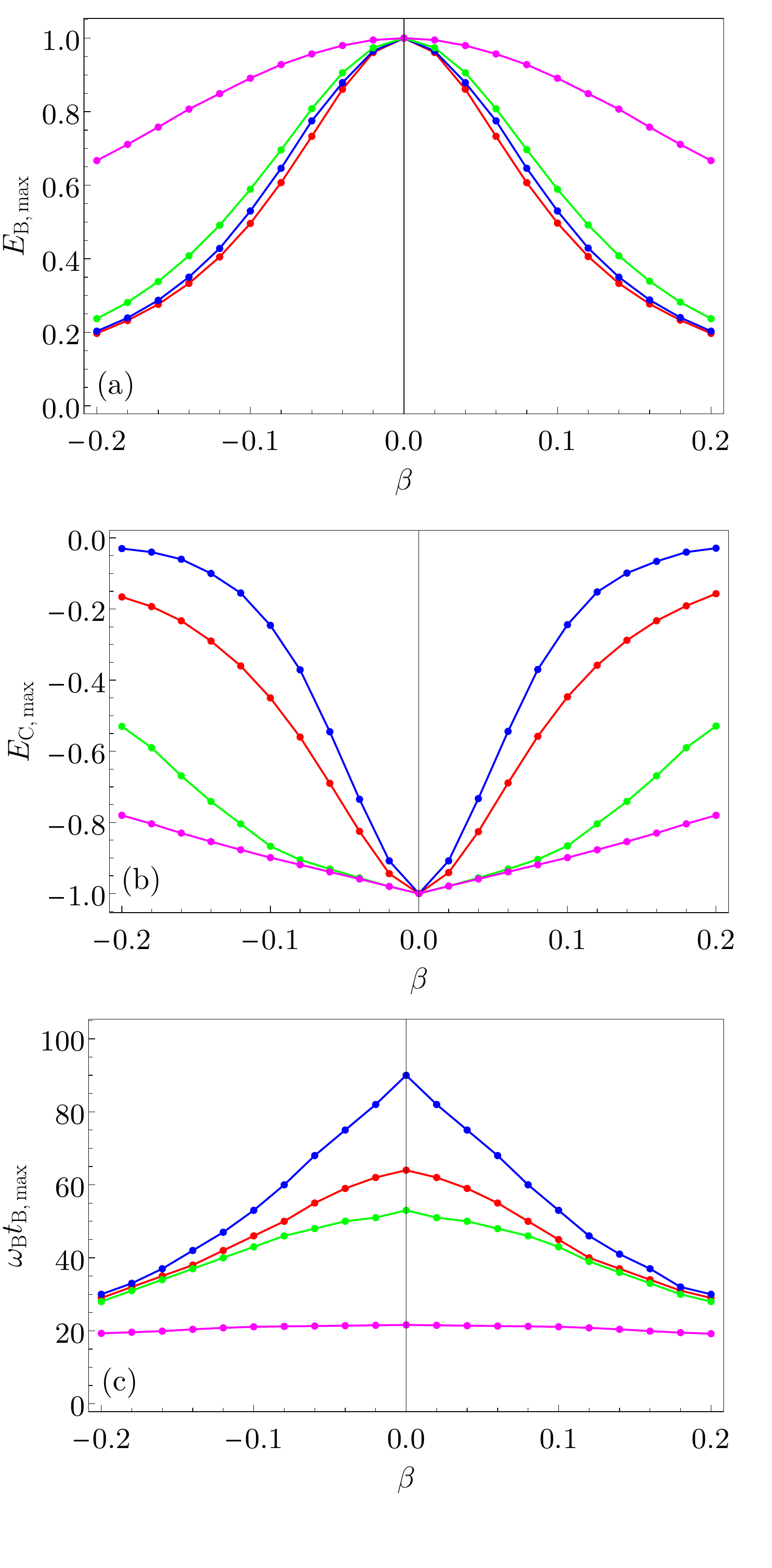} 
   \caption {(a) $E_{\rm B, max}$, (b) $E_{\rm C, max}$ (in units of $\omega_{\rm B}$) and $\omega_{\rm B}t_{\rm B,max}$ (c) as function of $\beta$ for the direct energy transfer (red dots), the TLS-mediated model (blue dots), the cavity-mediated model with $n=1$ (green dots) and the cavity-mediated model with $n=8$ (magenta dots). Other parameters are $g=0.05\omega_{\rm B}$ and $\omega_{\rm B}t_0=0.1$. Values of $\tau$ are chosen for each point in such a way to switch off the energy transfer in the system when the maximum energy transfer is achieved.}
   \label{fig8}
\end{figure}

Comparing Fig.~\ref{fig8} (a) and (b) we can see that for $\beta=0$ all the models allows a complete energy transfer from C to B. However, when we considered off-resonant regimes the cavity-mediated model, even in presence of a single photon, has the best performances in terms of energy transfer for all considered values of $\beta$. Moreover, we observe that a cavity with an high number of photons ($n=8$ for the magenta curve) is characterized by both a greater energy stored in B and a greater energy extracted C with respect to all the other considered devices even far from resonance.

The importance of a cavity mediated coupling also emerge when we consider the energy transfer time $t_{\rm B,max}$ [see Fig.~\ref{fig8} (c)] which is shorter with respect to the other considered models. In addition when we consider a cavity with a larger number of photons ($n=8$ for the considered case) we observe that this quantity is only marginally affected by the value of the parameter $\beta$. These aspects can play a role in situations where the average energy transfer power, namely the energy transferred in a given time, become the relevant parameter to judge the functionality of a device~\cite{Binder15,Campaioli17,Ferraro18, JuliaFarre20}. 

Moreover, the same qualitative behavior can be observed for different values of the coupling constant $g$, with the charging times of all the considered systems which scales as $g^{-1}$ as long as the RWA holds.

\section{Conclusions \label{sec:conclusions}}

We have investigated the coherent energy transfer between two TLSs mediated by an additional simple quantum system, namely a TLS or the photons in a resonant cavity. We have considered the TLSs sending and receiving energy both on and off-resonance and compared the energy transfer performances with the case of direct coupling. We have characterized various figures of merit, focusing in particular on the energy stored in the receiving TLS, initially empty, the energy extracted from the charger and the work done to switch off the interaction. We have shown that, while the TLS-mediated case shows no advantages with respect to the direct coupling case, in the cavity-mediated case the infinite Hilbert space of the harmonic oscillator can be exploited to improve the performance of the device. This analysis, together with the fact that the cavity-mediated interaction can lead to long distance connections between TLSs, can open new perspectives in the domain of quantum devices for energy storage and transfer, and more in general in the context of quantum technologies. 

Possible further developments of our study could address the robustness of the discussed phenomenology with respect to dissipation and interaction with an external environment. This will allow to identify the experimental platform more suitable for actual implementation of these devices.

\begin{acknowledgments}
This paper has been published thank to the funding of the European Union-NextGenerationEU through the "QUantum Busses for coherent EneRgy Transfer (QUBERT)" project, in the framework of the Curiosity Driven Grant 2021 of the University of Genova. Authors also acknowledge the financial support of the "Dipartimento di Eccellenza MIUR 2018-22". D. F. would like to thank M. Arzeo, P. Scarlino and M. Stern for useful discussions. 
\end{acknowledgments}


\appendix

\section{Capacitive coupling between superconducting circuits}\label{appA}

In this Appendix we consider simple superconducting circuits whose effective low energy descriptions map into the model Hamiltonians introduced in Section~\ref{sec:model} of the main text. For this analysis we will follow closely the derivation reported in Ref.~\cite{Krantz19}.

\subsection{Direct coupling between TLSs\label{sec:qubitmodels1}}

\begin{figure}[h!]
\centering
\includegraphics[scale=0.20]{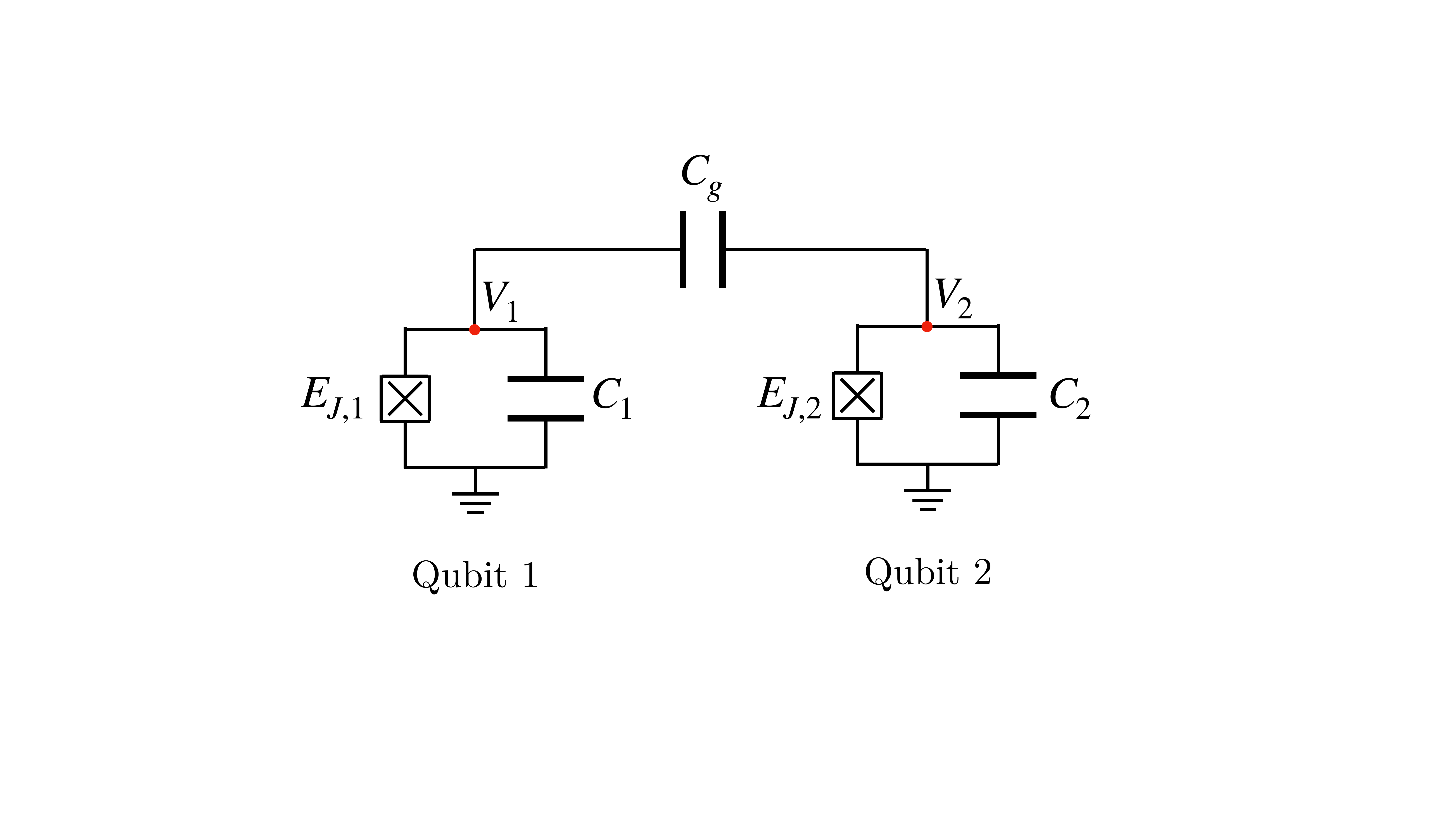} 
   \caption {Two superconducting circuits, with capacitance $C_{1/2}$ and Josephson energy $E_{J,1/2}$, play the role of qubits and are connected by means of the capacitance $C_g$. The crossed square symbols represent the Josephson junctions, while the voltages across each circuit are indicated with $V_{1/2}$.}
   \label{fig1a}
\end{figure}

Let's consider the scheme in Fig.~\ref{fig1a} where two superconducting circuits, each composed by a capacitor and a Josephson junction, are capacitively coupled through the capacitance $C_g$. 

The associated Hamiltonian can be written as

\begin{equation}
\label{Hqtot} 
H^{(\rm d)}=H_1+H_2+H_{\rm int}^{(\rm d)}, 
\end{equation}

\noindent where the Hamiltonians of the two separated circuits can be written as

\begin{equation}
\label{Hqubit} 
H_k=4 E_{C,k}n^2_k-E_{J,k}\cos \phi_k\qquad (k=1,2), 
\end{equation}

\noindent with 
\begin{equation}
E_{C,k}=\frac{e^2}{2(C_k+C_g)}
\end{equation} 
the charging energy of each circuit due to the capacitances $C_{k}$ and $C_{g}$ (see Fig.~(\ref{fig1a})) and $E_{J,k}$ the Josephson energy. With $n_k$ we indicate the Cooper pair number operator (with respect to a given background) and with $\phi_k$ the conjugate phase operator such that $[\phi_{k}, n_{k}]=i$. The direct capacitive coupling between the two circuits is described by the interaction Hamiltonian

\begin{equation} 
H_{\rm int}^{(\rm d)}= C_gV_1V_2, 
\end{equation}

\noindent with $V_{1/2}$ the voltage operators corresponding to the nodes in Fig.~(\ref{fig1a}). In the limit $C_g \ll C_{1,2}$ one can write

\begin{equation} 
H_{\rm int}^{(\rm d)}\simeq 4e^2\frac{C_g}{C_1C_2}n_1n_2.
\label{int_ham_d}
\end{equation}

\noindent Due to the cosine term in Eq.~(\ref{Hqubit}) we can then consider the circuits as coupled anharmonic oscillators. In the transmon regime~\cite{Koch07}, where the condition $E_{C,k}\ll E_{J,k}$ is satisfied, the free Hamiltonian in Eq.~(\ref{Hqubit}) can be rewritten (up to the fourth order in $\phi_k$) as an anharmonic oscillator of the Duffing type

\begin{equation} 
H_k\simeq 4E_{C,k}n_k^2+\frac{E_{J,k}}{2}\phi^2_k-\frac{E_{J,k}}{24}\phi^4_k. 
\label{H_Duffing}
\end{equation}
In terms of bosonic ladder operators such that

\begin{eqnarray}
n_k&=&i\bigg(\frac{E_{J,k}}{32E_{C,k}}\bigg)^{\frac{1}{4}}(b_k^\dagger-b_k)
\label{bk1}\\ 
\phi_k&=&\bigg(\frac{2E_{C,k}}{E_{J,k}}\bigg)^{\frac{1}{4}}(b_k^\dagger+b_k),
\label{bk2}
\end{eqnarray} 
the previous Hamiltonian becomes

\begin{equation} 
H_k\simeq \omega_k b_k^\dagger b_k-\alpha_k(b_k^\dagger+b_k)^4,
\label{H_b4}
\end{equation}

\noindent with $\omega_k=\sqrt{8E_{C,k}E_{J,k}}$ and $\alpha_k=E_{C,k}/12$.

Moreover, the interaction Hamiltonian in Eq.~(\ref{int_ham_d}) becomes

\begin{equation} 
H_{\rm int}^{(\rm d)}\simeq-g^{(\rm d)}(b_1^{\dagger}-b_{1})(b_2^{\dagger}-b_{2}), 
\end{equation}

\noindent where we have introduced the coupling constant

\begin{equation} 
g^{(\rm d)}=\frac{2 \sqrt{2}}{e^{2}}C_gE_{C,1}^{\frac{3}{4}}E_{C,2}^{\frac{3}{4}}E_{J,1}^{\frac{1}{4}}E_{J,2}^{\frac{1}{4}}, 
\end{equation}

\noindent which only depends on the physical parameters of the superconducting circuits.

Due to the anharmonicity of the energy levels, it is possible to focus only on the ground and the first excited state of each superconducting circuit, neglecting the other excited states. Then, introducing a more convenient spin operator notation to describe these effective TLSs (qubits), it is possible to rewrite the initial Hamiltonian in Eq.~(\ref{H_b4}), up to a constant, as

\begin{equation} 
H^{(\rm d)}\simeq \sum_{k=1,2} \frac{\omega_k}{2}\sigma_{z,k}+g^{(\rm d)}\sigma_{y,1}\sigma_{y,2}, 
\label{H_sigmay}
\end{equation}

\noindent which is the Hamiltonian of two $1/2$ spins coupled through exchange interaction. In the framework of the RWA~\cite{Schleich_Book}, where $g^{(\rm d)}\ll \omega_{1}, \omega_{2}$ and the two level spacing are not far from resonance, the counter-rotating terms only play a minor role and can be neglected. Under these assumptions, Eq.~(\ref{H_sigmay}) can be finally rewritten as

\begin{equation} 
H_{\rm RWA}^{\rm(d)}=\sum_{k=1,2} \frac{\omega_k}{2}\sigma_{z,k}+g^{(\rm d)}(\sigma_{-,1}\sigma_{+,2}+\sigma_{+,1}\sigma_{-,2}), 
\end{equation}
which, once considered the possibility to switch on and off the interaction, corresponds to the expression introduced in the main text for the direct coupling between two TLSs ($H_{\rm TLS}^{\rm (d)}$) as well as the main building block needed to realize a chain of three TLSs with local interaction ($H_{\rm TLS}^{\rm (m)}$).

\subsection{Cavity mediated coupling\label{sec:qubitmodels2}}

\begin{figure}[h!]
\centering
\includegraphics[scale=0.14]{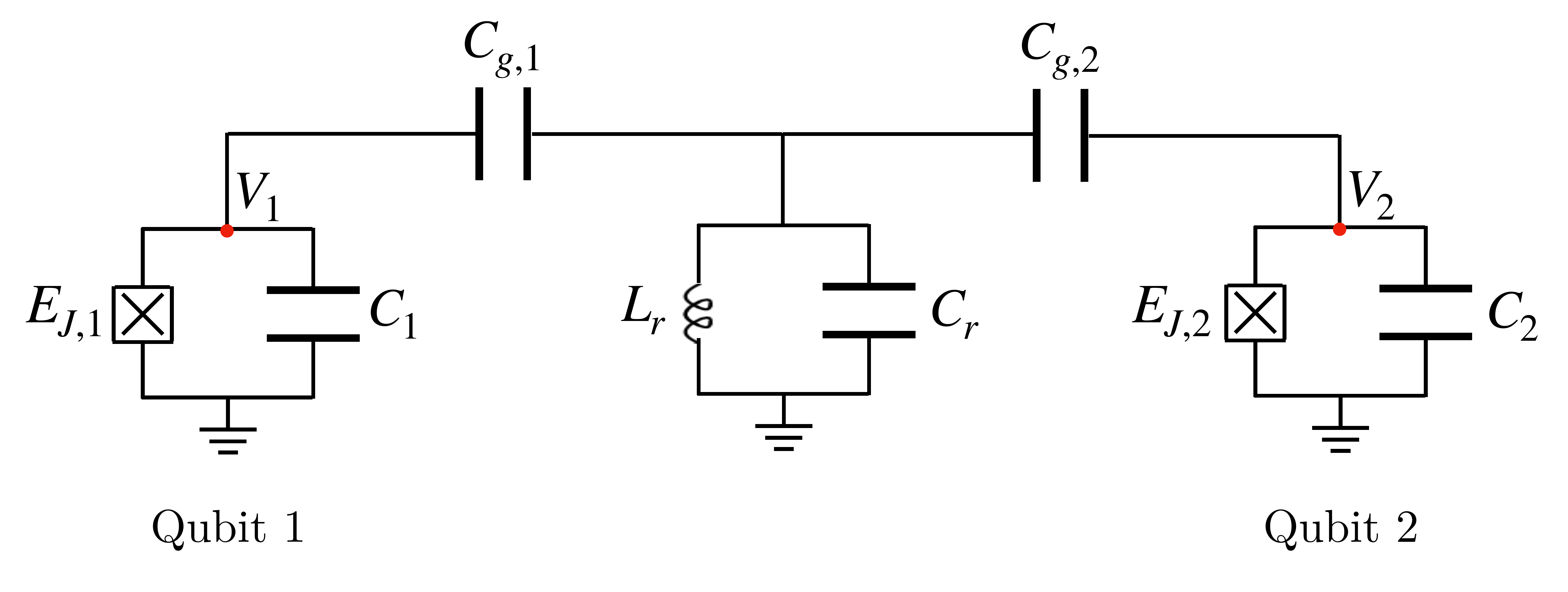} 
   \caption {Two superconducting circuits, with capacitance $C_{1/2}$ and Josephson energy $E_{J, 1/2}$, are connected through a LC resonator with capacitance $C_r$ and inductance $L_r$. The crossed square symbols represent the Josephson junctions, while the voltages across each circuit are indicated by $V_{1/2}$.}
   \label{fig2a}
\end{figure}

Let's consider now the scheme in Fig.~\ref{fig2a}, where two superconducting circuits are coupled through a LC resonator with capacitance $C_r$ and inductance $L_r$, which play the role of a cavity. 

In analogy to what done above we can write the Hamiltonian of the system as

\begin{equation}
\label{Hqm} 
H^{(\rm m)}=\sum_{k=1,2} 4E_{{C,k}}(n_k+n_r)^2-E_{J,k}\cos\phi_k + \omega_r a^\dagger a, 
\end{equation}

\noindent where $\omega_r=1/\sqrt{C_r L_r}$ the frequency of the harmonic oscillator describing the LC circuit and 

\begin{equation}
n_r=\frac{i}{2e}\frac{C_g}{C_r}\left(\frac{C_r}{4L_r}\right)^{\frac{1}{4}}(a^{\dagger}-a), 
\end{equation}

\noindent with $a$ ($a^\dagger$) the annihilation (creation) operator for the photons in the resonant cavity.

Taking into account again the transmon limit in Eq.~(\ref{H_Duffing}) and in terms of the $b_k$ operators introduced in Eqs.~(\ref{bk1}) and (\ref{bk2}) the Hamiltonian in Eq.~(\ref{Hqm}) can be rewritten as

\begin{eqnarray} 
H^{(\rm m)}&\simeq&\sum_{k=1,2} [\omega_kb_k^\dagger b_k-\alpha_k(b^\dagger_k+b_k)^4\nonumber\\ 
&-&g_k^{(\rm m)}(b_k^\dagger-b_k)(a^\dagger-a)] + \omega_ra^\dagger a, 
\end{eqnarray}

\noindent where we have considered a renormalization of the LC frequency $\omega_{r}$ due to the $n_r^2$ term and we have introduced the effective dipole coupling between matter and radiation

\begin{equation} 
g_k^{(\rm m)}=\frac{2^{\frac{1}{4}}}{e}E^{\frac{3}{4}}_{C,k}E^{\frac{1}{4}}_{J, k}\frac{C_g}{C_r}\bigg(\frac{C_r}{4L_r}\bigg)^{\frac{1}{4}}. 
\end{equation}

Considering again the anharmonicity of the qubits and performing the RWA one obtains, up to a constant,

\begin{equation} H_{\rm RWA}^{(\rm m)}=\sum_{k=1,2} \bigg[\frac{\omega_k}{2}\sigma_{z,k}+g_k^{(\rm m)}(a^\dagger\sigma_{-,k}+a\sigma_{-k})\bigg]+\omega_ra^\dagger a, \end{equation}
which, once considered the possibility to switch on and off the interaction, leads to the cavity mediated interaction between TLSs discussed in the main text ($H_{\rm cavity}^{(\rm m)}$).


\section{Considerations about power and work \label{sec:dimwork}}

Here we demonstrate that, after switching off the interaction, the work done on the three different models discussed in the main text has the same expression.

We start by recalling and further specifying the definition of power given in the main text, namely  
\begin{eqnarray} 
\label{Ptrace}
P(t)&\equiv& \frac{d}{dt}\left[{\rm Tr}\{\rho(t) H(t)\}\right]\nonumber\\ 
&=& {\rm Tr}\{\rho(t)\left[H(t), H(t) \right]\} +{\rm Tr}\bigg\{\rho(t)\frac{\partial H(t)}{\partial t}\bigg\}.\nonumber\\
\label{H_expl}
\end{eqnarray}
Notice that the commutator in the last line is obviously zero, however we have written it explicitly in order to better clarify the following considerations.

In the subsection below, we will investigate this quantity for the various discussed cases connecting it to the work needed to switch on and off the interaction. 

\subsection{TLS-TLS model}

In the case of a direct energy transfer between two TLSs the total Hamiltonian of the system [indicated with $H(t)$ in Eq.~(\ref{H_expl})] is given by the sum of Eq.~(\ref{H_0_d}) and Eq.~(\ref{H_int_d}). According to this, it is possible to rearrange Eq.~(\ref{H_expl}) in such a way that

\begin{widetext}

\begin{eqnarray} 
P(t)&=&\frac{d E_{\rm C}}{dt}+\frac{d E_{\rm B}}{dt}+\frac{d E_{\rm int}}{dt} \nonumber \\
&=&igf(t)(1-\alpha)\omega_{\rm B}{\rm Tr}\{\rho(t)(\sigma_-^{\rm(C)}\sigma_+^{\rm(B)}-\sigma_+^{\rm(C)}\sigma_-^{\rm(B)})\}+\frac{d E_{\rm int}}{dt}\nonumber \\
\label{power_TLS} &=&(1-\alpha)\frac{d E_{\rm B}}{dt}+\frac{d E_{\rm int}}{dt},
\label{P_d}
\end{eqnarray}

\end{widetext}

\noindent where we have used the fact that $\omega_{\rm C}=\alpha \omega_{\rm B}$, as in Eq.~(\ref{offR}).
Now, integrating Eq.~(\ref{power_TLS}) between times $0$ and $t$, taking into account the definitions and the initial conditions discussed in Section \ref{Sec_IIIA} we obtain the expression for the work at a given time $t$

\begin{equation}
W(t)=\int^{t}_{0} dt' P(t')=(1-\alpha)E_{\rm B}(t)+ E_{\rm int}(t).
\end{equation}

\begin{figure}[h!]
\centering
\includegraphics[scale=0.5]{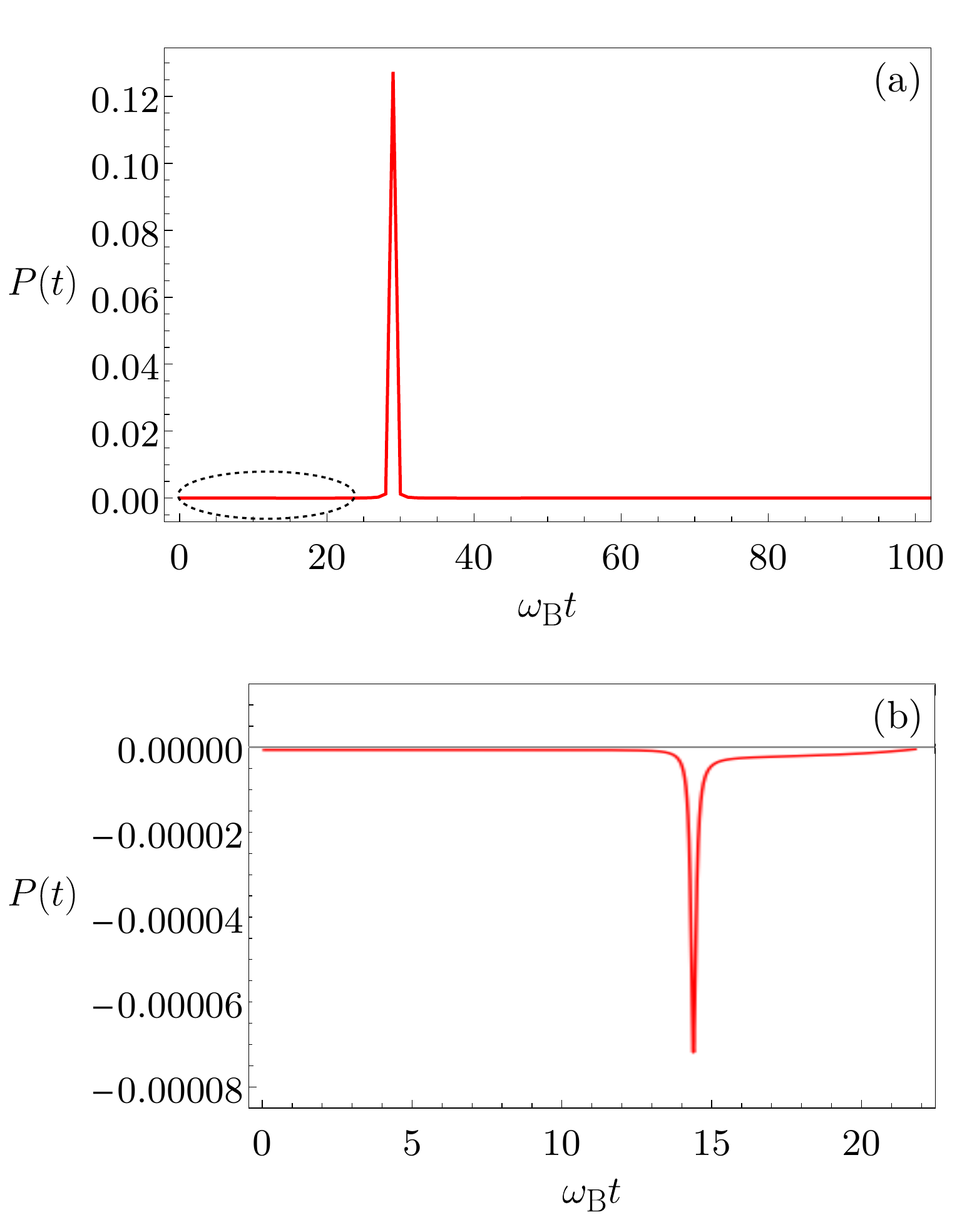} 
   \caption {Behaviour of the power $P(t)$ (in units of $\omega_{\rm B}^2$) as function of $\omega_{\rm B}t$ for the direct-coupling case at $\alpha=0.8$. The inset shows a zoom of the curve in correspondence of the injection time [dashed circle in panel (a)]. Other parameters are the ones in Fig.~\ref{fig2b} of the main text.}
   \label{figA3}
\end{figure}

If we consider a time $t^\star> \tau$ such that the interaction is turned off and the energy stored in the battery has reached its maximum $E_{\rm B, max}$, the work reduces to

\begin{equation}
\label{Wcost} 
W^\star\equiv W(t^\star)=(1-\alpha)E_{\rm B,max}. 
\end{equation}

Notice that for such times the work done on the system is constant and only depends on the maximum value of the energy stored in the QB and on the mismatch $\alpha$ in the level spacing between the TLSs.

To better understand the previous discussion, in Fig.~\ref{figA3} we report the power, as in Eq.~(\ref{P_d}), expected for an off-resonant case. As we can see, this quantity is zero everywhere, except for a very small negative peak in correspondence of the switching on of the interaction (see inset) and a more pronounced peak when the interaction is switched off. This peaked structure is a direct consequence of the functional behavior of the time derivative of the function $f(t)$, hidden in the last term of Eq.~(\ref{P_d}), while the intensity of the peaks is related to the state of the system at the considered time. Indeed, the first peak is closed to be zero due to the considered initial conditions and becomes null in the case of an infinitely sharp switching on of the interaction ($\omega_{B}t_{0}\rightarrow 0$). Conversely the second peaks reduces to zero only when the QB is perfectly charged, situation which is not achieved in the off-resonant case.

\subsection{TLS-mediated model}

For a system composed by three TLSs $H(t)$ in Eq.~(\ref{H_expl}) is given by Eq.~(\ref{H3tls}). According to this and properly rearranging the terms in Eq.~(\ref{H_expl}) we can write the power as

\begin{widetext}

\begin{eqnarray}  
P(t)&=&\frac{d E_{\rm C}}{dt}+\frac{d E_{\rm B}}{dt}+\frac{d E_{\rm M}}{dt}+\frac{d E_{\rm int}}{dt}\nonumber\\
&=&igf(t)(1-\alpha)\omega_{\rm B}{\rm Tr}\{\rho(t)(\sigma_+^{\rm(B)}\sigma_-^{\rm(M)}-\sigma_-^{\rm(B)}\sigma_+^{\rm(M)})\}+\frac{d E_{\rm int}}{dt} \nonumber \\
&=&(1-\alpha)\frac{d E_{\rm B}}{dt}+\frac{d E_{\rm int}}{dt},
\end{eqnarray}

\end{widetext}

\noindent where we have considered $\omega_{\rm C}=\omega_{\rm M}=\alpha \omega_{\rm B}$. The power just obtained naturally leads to the same expression of $W^{*}$ derived in Eq.~(\ref{Wcost}) for times $t^\star>\tau$. Moreover, the same qualitative behaviour discussed above for the power as a function of time is obtained in this case. Therefore, in order not to be pedantic we don't report the associated figure for the present model.

\subsection{Cavity-mediated model}

In the case of a cavity-mediated interaction between TLSs $H(t)$ in Eq.~(\ref{H_expl}) is given by Eq.~(\ref{Hcav}). According to this and properly rearranging the terms in Eq.~(\ref{H_expl}) the power can be written as

\begin{widetext}

\begin{eqnarray} 
P(t)&=&\frac{d E_{\rm C}}{dt}+\frac{d E_{\rm B}}{dt}+\frac{d E_{\rm M}}{dt}+\frac{d E_{\rm int}}{dt}\nonumber\\
&=&igf(t)(1-\alpha)\omega_{\rm B}{\rm Tr}\{\rho(t)(a^\dagger\sigma_-^{(\rm B)}-a\sigma_+^{\rm(B)})\}+\frac{d E_{\rm int}}{dt} \nonumber \\
\label{power} &=&(1-\alpha)\frac{d E_{\rm B}}{dt}+\frac{d E_{\rm int}}{dt},
 \end{eqnarray}
 
 \end{widetext}

\noindent where we have again taken into account the condition $\omega_{\rm C}=\omega_{\rm M}=\alpha \omega_{\rm B}$ in Eq.~(\ref{offR}). Also in this case, after integration, we obtains the same expression of $W^{*}$ derived in Eq.~(\ref{Wcost}) for times $t^\star>\tau$. Moreover, the qualitative behaviour of the power is again analogous to the one depicted in Fig.~\ref{figA3}.

\section{Conserved quantities and constraints \label{sym}}

To solve the dynamics of the discussed models we resort to the exact numerical diagonalization of the corresponding Hamiltonians. In all cases the dimensions of the Hilbert spaces are constrained by overall conservation laws valid in the RWA, leading to a simplification of the analysis. Before addressing separately the three cases discussed in the main text it is useful to recall the identities for commutators $[\sigma_z,\sigma_\pm]=\pm2\sigma_\pm$; $\left[\sigma_+,\sigma_-\right]=\sigma_z$.

\subsection{Direct coupling between TLSs \label{sym1}}

In the case of the direct energy transfer between two TLSs one has that the operator 

\begin{eqnarray} N_{\rm spin}^{\rm (d)}&\equiv& \sigma_+^{(\rm C)}\sigma_-^{(\rm C)}+\sigma_+^{(\rm B)}\sigma_-^{(\rm B)} \nonumber \\
&=&\frac{1}{2}\bigg(\sigma_z^{\rm (C)}+\sigma_z^{\rm (B)}\bigg)+1, \end{eqnarray}
related to the $z$ component of the total spin of the system, satisfies the commutation relation 

\begin{equation} [H_{\rm TLS}^{\rm (d)}(t), N_{\rm spin}^{\rm (d)}]=[H_{\rm int}^{\rm (d)}(t), N_{\rm spin}^{\rm (d)}]=0. \end{equation}

This implies that $N_{\rm spin}^{\rm (d)}$ is a conserved quantum number which can be exploited to constraint the Hilbert space and consequently the dimension of the matrix Hamiltonian. Starting from the initial state $|\psi(0)\rangle=|1_{\rm C}, 0_{\rm B}\rangle$ (with eigenvalue of $N_{\rm spin}^{\rm (d)}$ equal to one), the dynamics of the system is constrained in the two-dimensional subspace spanned by the vector states

\begin{equation} 
|\varphi_1\rangle=|1_{\rm C}, 0_{\rm B}\rangle \quad\quad |\varphi_2\rangle= |0_{\rm C}, 1_{\rm B}\rangle. 
\end{equation}
Consequently we need to diagonalize a $2\times2$ matrix Hamiltonian of the form

\begin{equation} 
H_{\rm TLS}^{\rm (d)}(t)=\begin{pmatrix}\dfrac{\omega_{\rm C}-\omega_{\rm B}}{2} &gf(t) \\ gf(t)& -\dfrac{\omega_{\rm C}-\omega_{\rm B}}{2}\end{pmatrix}. 
\end{equation}

\subsection{TLS-mediated case \label{sym2}}

In this case, extending what done before, we can introduce the operator

\begin{eqnarray} 
N_{\rm spin}^{\rm (m)}&\equiv& \sigma_+^{(\rm C)}\sigma_-^{(\rm C)}+\sigma_+^{(\rm B)}\sigma_-^{(\rm B)}+\sigma_+^{(\rm M)}\sigma_-^{(\rm M)} \nonumber \\
&=&\frac{1}{2}\bigg(\sigma_z^{\rm (C)}+\sigma_z^{\rm (B)}+\sigma_z^{\rm (M)}\bigg)+\frac{3}{2} 
\end{eqnarray}
related again to the $z$ component of the total spin of the system.

We then can evaluate the commutator

\begin{eqnarray}  [H_{\rm TLS}^{\rm (m)}(t), N_{\rm spin}^{\rm (m)}]&=&gf(t)[\sigma_-^{(\rm C)}\sigma_+^{(\rm M)}+\sigma_+^{(\rm C)}\sigma_-^{(\rm M)}, N_{\rm spin}^{\rm (m)}] \nonumber \\
&+&gf(t)[\sigma_-^{(\rm B)}\sigma_+^{(\rm M)}+\sigma_+^{(\rm B)}\sigma_-^{(\rm M)}, N_{\rm spin}^{\rm (m)}]=0. \nonumber \\
 \end{eqnarray}

Consequently $N_{\rm spin}^{\rm (m)}$ is a conserved quantum number which can be exploited to constraint the Hilbert space. If the initial state of the system is in an arbitrary superposition for the mediating TLS
\begin{equation}
|\psi (t)\rangle=|1_{\rm C}, 0_{\rm B}\rangle \otimes (a |0_{\rm M}\rangle + b |1_{\rm M}\rangle),
\end{equation}
with $a,b \in \mathbb{C}$ such that $|a|^{2}+|b|^{2}=1$, then the evolution of the system is limited to the space spanned by the six states vectors with eigenvalues of $N_{\rm spin}^{\rm (m)}$ equal to one or two, namely  

\begin{eqnarray}
|\varphi_1\rangle=|1_{\rm C}, 0_{\rm B}, 0_{\rm M} \rangle &\quad& |\varphi_2\rangle=|0_{\rm C}, 1_{\rm B}, 0_{\rm M} \rangle \nonumber\\ 
|\varphi_3\rangle=|0_{\rm C}, 0_{\rm B}, 1_{\rm M} \rangle &\quad& |\varphi_4\rangle=|1_{\rm C}, 0_{\rm B}, 1_{\rm M} \rangle \nonumber\\ 
|\varphi_5\rangle=|0_{\rm C}, 1_{\rm B}, 1_{\rm M} \rangle &\quad& |\varphi_6\rangle=|1_{\rm C}, 1_{\rm B}, 1_{\rm M} \rangle. 
\end{eqnarray}
According to the above considerations we need to diagonalize a $6 \times 6$ matrix Hamiltonian of the form

\begin{widetext}

\begin{equation}
H_{\rm TLS}^{\rm (m)}(t)=\begin{pmatrix}
0 & 0 & 0 & \frac{\omega_{\rm C}-\omega_{\rm B}-\omega_{\rm M}}{2} & 0 & gf(t) \\
0 & 0 & 0 & 0 & \frac{-\omega_{\rm C}+\omega_{\rm B}-\omega_{\rm M}}{2} & gf(t) \\
0 & 0 & 0 & gf(t) & gf(t) & \frac{-\omega_{\rm C}-\omega_{\rm B}+\omega_{\rm M}}{2} \\
gf(t) & \frac{\omega_{\rm C}-\omega_{\rm B}+\omega_{\rm M}}{2} & 0 & 0 & 0 & 0 \\
gf(t) & 0 & \frac{-\omega_{\rm C}+\omega_{\rm B}+\omega_{\rm M}}{2} & 0 & 0 & 0 \\
\frac{\omega_{\rm C}+\omega_{\rm B}-\omega_{\rm M}}{2} & gf(t) & gf(t) & 0 & 0 & 0 
\end{pmatrix}.
\end{equation}

\end{widetext}

\subsection{Cavity-mediated coupling \label{sym3}}

In the cavity-mediated case it is possible to define the excitations number operator~\cite{Schleich_Book}

\begin{eqnarray} \xi^{\rm (m)}&\equiv& \sigma_+^{(\rm C)}\sigma_-^{(\rm C)}+\sigma_+^{(\rm B)}\sigma_-^{(\rm B)}+a^\dagger a \nonumber \\
&=&\frac{1}{2}\bigg(\sigma_z^{\rm (C)}+\sigma_z^{\rm (B)}\bigg)+a^\dagger a+1. \end{eqnarray}

One can easily verify that 

\begin{eqnarray} [H_{\rm cavity}^{\rm (m)}(t), \xi^{\rm (m)}]&=&gf(t)[a^\dagger \sigma_-^{(\rm C)}+a \sigma_+^{(\rm C)}, \xi^{\rm (m)}] \nonumber \\
&+&gf(t)[a^\dagger \sigma_-^{(\rm B)}+a \sigma_+^{(\rm B)}, \xi^{\rm (m)}]=0. \nonumber \\
 \end{eqnarray}

In this case $\xi^{\rm (m)}$ is a conserved quantum number for the cavity-mediated system and consequently we can exploit it to constrain the Hilbert space. Starting from the initial state 
\begin{equation}
|\psi(t)\rangle=|1_{\rm C}, 0_{\rm B}\rangle \otimes (a |n\rangle + b |n+1\rangle),  
\end{equation} 
with $n$ number of photons and $a,b \in \mathbb{C}$ such that $|a|^{2}+|b|^{2}=1$, the dynamics of the system is limited to the space spanned by the eight state vectors with eigenvalue of $\xi^{\rm (m)}$ given by $n+1$ and $n+2$, namely 

\begin{eqnarray}
|\varphi_1\rangle&=&|1_{\rm C}, 1_{\rm B}, n-1 \rangle \quad\quad |\varphi_2\rangle=|1_{\rm C}, 0_{\rm B}, n \rangle \nonumber\\ 
|\varphi_3\rangle&=&|0_{\rm C}, 1_{\rm B}, n \rangle \quad\quad\quad \  \ |\varphi_4\rangle=|0_{\rm C}, 0_{\rm B}, n+1 \rangle \nonumber\\ |\varphi_5\rangle&=&|1_{\rm C}, 1_{\rm B}, n \rangle \quad\quad\quad \  \  |\varphi_6\rangle=|1_{\rm C}, 0_{\rm B}, n+1 \rangle  \nonumber \\
|\varphi_7\rangle&=&|0_{\rm C}, 1_{\rm B}, n+1 \rangle \quad\quad |\varphi_8\rangle=|0_{\rm C}, 0_{\rm B}, n+2 \rangle. \nonumber \\
\end{eqnarray}
We then need to diagonalize the $8\times 8$ matrix Hamiltonian of the form

\begin{widetext}

\begin{equation}
\scriptstyle H_{\rm cavity}^{\rm (m)}(t)=\begin{pmatrix}
\scriptstyle \frac{\omega_{\rm C}+\omega_{\rm B}}{2}+m\omega_{\rm M} & \scriptstyle g(t)\sqrt{n} & \scriptstyle g(t)\sqrt{n} &\scriptstyle 0 &\scriptstyle 0 &\scriptstyle 0 &\scriptstyle 0 &\scriptstyle 0 \\
\scriptstyle g(t)\sqrt{n} & \scriptstyle \frac{\omega_{\rm C}-\omega_{\rm B}}{2}+n\omega_{\rm M} & \scriptstyle 0 & \scriptstyle g(t)\sqrt{p} &\scriptstyle 0 &\scriptstyle 0 &\scriptstyle 0 &\scriptstyle 0 \\
\scriptstyle g(t)\sqrt{n} &\scriptstyle 0 &\scriptstyle \frac{-\omega_{\rm C}+\omega_{\rm B}}{2}+n\omega_{\rm M} &\scriptstyle g(t)\sqrt{p} &\scriptstyle 0 &\scriptstyle 0 &\scriptstyle 0 &\scriptstyle 0 \\
\scriptstyle0 & \scriptstyle g(t)\sqrt{p} & \scriptstyle g(t)\sqrt{p} & \scriptstyle \frac{-\omega_{\rm C}-\omega_{\rm B}}{2}+p\omega_{\rm M} &\scriptstyle 0 &\scriptstyle 0 &\scriptstyle 0 &\scriptstyle 0 \\
\scriptstyle0 &\scriptstyle 0 &\scriptstyle 0 &\scriptstyle 0 & \scriptstyle \frac{\omega_{\rm C}+\omega_{\rm B}}{2}+n\omega_{\rm M} & \scriptstyle g(t)\sqrt{p} & \scriptstyle g(t)\sqrt{p} &\scriptstyle 0 \\
\scriptstyle0 &\scriptstyle 0 &\scriptstyle 0 &\scriptstyle 0 & \scriptstyle g(t)\sqrt{p} & \scriptstyle \frac{\omega_{\rm C}-\omega_{\rm B}}{2}+p\omega_{\rm M} &\scriptstyle 0 & \scriptstyle g(t)\sqrt{q} \\
\scriptstyle0 &\scriptstyle 0 &\scriptstyle 0 &\scriptstyle 0 & \scriptstyle g(t)\sqrt{p} &\scriptstyle 0 & \scriptstyle \frac{-\omega_{\rm C}+\omega_{\rm B}}{2}+p\omega_{\rm M} & \scriptstyle g(t)\sqrt{q} \\
\scriptstyle0 &\scriptstyle 0 &\scriptstyle 0 &\scriptstyle 0 &\scriptstyle 0 & \scriptstyle g(t)\sqrt{q} & \scriptstyle g(t)\sqrt{q} & \scriptstyle \frac{-\omega_{\rm C}-\omega_{\rm B}}{2}+q\omega_{\rm M}
\end{pmatrix},
\end{equation}
where we have introduced the short notation $g(t)=gf(t)$, $m=n-1$, $p=n+1$ and $q=n+2$.

\end{widetext}


\end{document}